\documentclass[fp,twocolumn]{jpsj3}
\usepackage{txfonts}
\usepackage{graphicx}
\usepackage{dcolumn}
\usepackage{bm}
\usepackage[dvipsnames]{xcolor}

\title{Non-Collinear and Non-Coplanar Magnetic Orders in 1/1 Periodic Approximant to the Icosahedral Quasicrystal}

\author{Shinji Watanabe$^{*}$ and Tatsuya Iwasaki}
\inst{Department of Basic Sciences, Kyushu Institute of Technology, Kitakyushu, Fukuoka 804-8550, Japan}

\abst{
Ground-state properties of rare-earth based 1/1 periodic approximants of icosahedral quasicrystal are clarified theoretically on the basis of an effective model for magnetism taking into account uniaxial anisotropy arising from crystalline electric field. By performing numerically-exact calculation on the 1/1 approximant crystal with a lattice constant $a=14.725~{\rm \AA}$, we have determined the ground-state phase diagram for ferromagnetic interactions. The result shows that eight kinds of noncollinear and noncoplanar magnetic structures are stabilized, whose magnetic space groups are identified as $I_{\rm P}m'\bar{3}'$, $C2'/m'$ and $R\bar{3}$. We have clarified the degeneracy of each ground state, which is expected to be reflected in the numbers of the domains. By analyzing each state, the magnetic as well as topological properties are revealed. Our results are shown to explain the measured magnetic structures in the 1/1 approximants and the effective model is discussed to be useful for understanding the magnetic structures and their topological properties in broad range of rare-earth based 1/1 approximants.
}

\begin{document}

\def\gsim{\mathop {\vtop {\ialign {##\crcr 
$\hfil \displaystyle {>}\hfil $\crcr \noalign {\kern1pt \nointerlineskip } 
$\,\sim$ \crcr \noalign {\kern1pt}}}}\limits}
\def\lsim{\mathop {\vtop {\ialign {##\crcr 
$\hfil \displaystyle {<}\hfil $\crcr \noalign {\kern1pt \nointerlineskip } 
$\,\,\sim$ \crcr \noalign {\kern1pt}}}}\limits}

\maketitle

\section{Introduction}

Quasicrystals (QCs) have no lattice periodicity but possess long-range structural order. 
Since the discovery of QC~\cite{Shechtman}, understanding of lattice structures has proceeded~\cite{Tsai,Takakura}. However, the electronic property is far from complete understanding because Bloch theorem based on translational invariance can no longer be applied. 

One of the unresolved issues on QCs has been whether magnetic long-range order is realized in three-dimensional QCs. 
To explore magnetic long-range order in QCs, experimental efforts have been devoted~\cite{Fisher1999,Sato2000,Goldman2013,Das2023}. 
There exist periodic crystals with local atomic configuration common to that in QC, which is referred to as approximant crystal (AC). 
Rare-earth based QCs and ACs have attracted much interest as a new research field of the strongly correlated electron system~\cite{NKSato2022}, which offer promising 
platform  
to search for the magnetic long-range order. 

\begin{figure}[b]
\includegraphics[width=8cm]{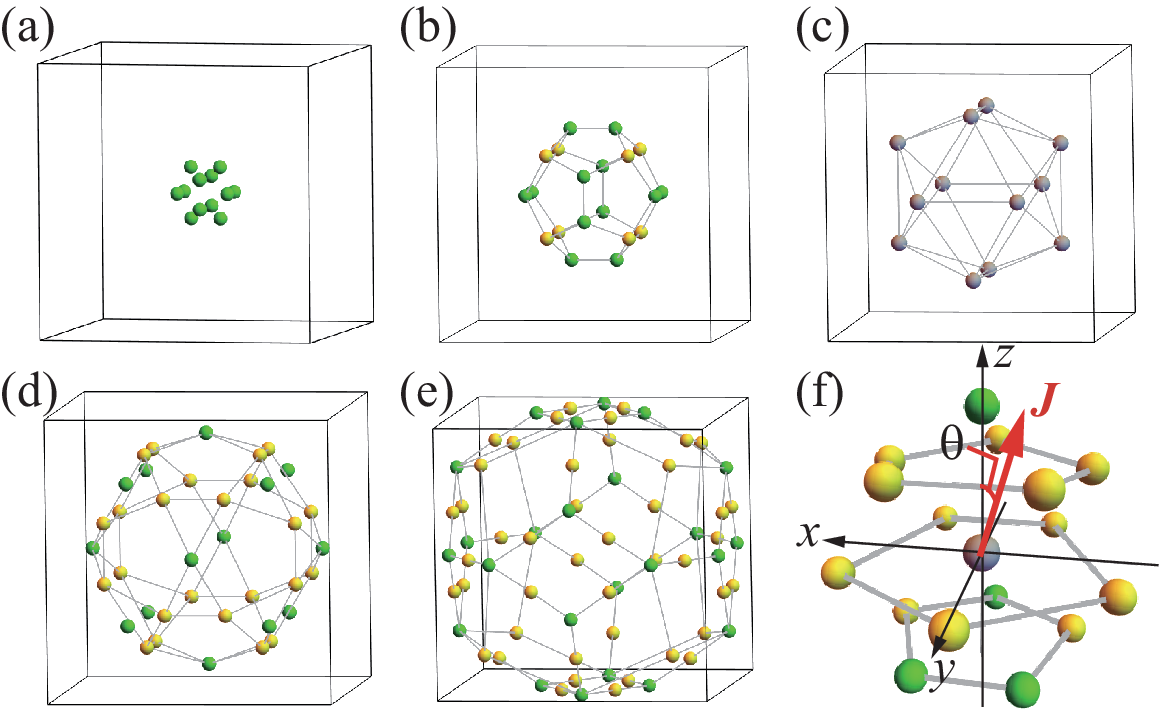}
\caption{(Color online) 
Tsai-type cluster of the Au-SM-Tb consists of (a) cluster center, (b) dodecahedron, (c) icosahedron, (d) icosidodecahedron, and (e) defect rhombic triacontahedron with Au (yellow), Au/SM (green), and Tb (gray). 
The frame box is the unit cell of the (expanded) unit cell of the 1/1 AC Au-SM-Tb. 
(f) The local configuration of atoms surrounding the Tb atom in the Tsai-type cluster. 
The angle $\theta$ is defined between the magnetic easy axis ${\bm J}$ for the CEF ground state and the pseudo 5-fold axis ($z$ axis). 
}
\label{fig:Tb_local}
\end{figure}

A class of the rare-earth based QCs and ACs is composed of the Tsai-type cluster, which consists of concentric shell structures of atoms as 
the cluster center [Fig.~\ref{fig:Tb_local}(a)], the dodecahedron [Fig.~\ref{fig:Tb_local}(b)], the icosahedron [Fig.~\ref{fig:Tb_local}(c)], the icosidodecahedron [Fig.~\ref{fig:Tb_local}(d)], and the defect rhombic triacontahedron [Fig.~\ref{fig:Tb_local}(e)]. 
The rare-earth atoms are located at 12 vertices of the icosahedron [see Fig.~\ref{fig:Tb_local}(c)]. 
The 4f electrons at the rare-earth atoms are responsible for the magnetism. 
In the 1/1 AC, there are two icosahedrons in the (expanded) unit cell of the body-center-cubic (bcc) lattice which is illustrated as the frame box in Figs.~\ref{fig:Tb_local}(a)-\ref{fig:Tb_local}(e).

In the rare-earth based 1/1 ACs, magnetic long-range orders have been observed by bulk measurements such as the specific heat and the magnetic susceptibility~\cite{Suzuki2021}. 
The antiferromagnetic (AFM) order has been observed in the 1/1 AC Cd$_6$R (R=Tb, Y, Pr, Nd, Sm, Gd, Tb, Dy, Ho, Er, Tm, Yb, and Lu) ~\cite{Tamura2010,Mori} and in the 1/1 AC Au-Al-R (R=Gd and Tb)~\cite{Ishikawa2018,Labib2025}. 
The ferromagnetic (FM) order has been observed 
in the 1/1 AC Au-SM-R (SM=Si, Ge, and Al; R=Gd, Tb, Dy, and Ho)~\cite{Hiroto2013,Hiroto2014,Ishikawa2016,Shiino2022}. 

Recently, ferromagnetic (FM) long-rage order has been discovered in QC Au$_{65}$Ga$_{20}$R$_{15}$ (R=Gd and Tb)~\cite{Tamura2021} and Au$_x$Ga$_{85-x}$Dy$_{15}$ ($x=62$-68)~\cite{Takeuchi2023}. 
Very recently, AFM long-range order has been discovered in QC Au$_{56}$In$_{28.5}$Eu$_{15.5}$~\cite{Tamura2025}. 
These findings have attracted great interest. 

Among the rare-earth-based ACs, magnetic structures have recently been identified by neutron measurements in 1/1 ACs Au$_{72}$Al$_{14}$Tb$_{14}$~\cite{Sato2019}, Au$_{70}$Si$_{17}$Tb$_{13}$~\cite{Hiroto}, Au-Si-R (R=Tb, Ho)~\cite{Gebresenbut}, and Au$_{65}$Ga$_{21}$Tb$_{14}$~\cite{Nawa2023}, where the non-collinear and non-coplanar magnetic structures have been revealed. 
In the 1/1 ACs Au$_{72}$Al$_{14}$Tb$_{14}$~\cite{Sato2019} and Au$_{65}$Ga$_{21}$Tb$_{14}$~\cite{Nawa2023}, the antiferromagnetic distribution of the magnetic states on the icosahedrons at the center and the corner of the unit cell is realized, while in the 1/1 ACs Au$_{70}$Si$_{17}$Tb$_{13}$~\cite{Hiroto} and Au-Si-R (R=Tb, Ho)~\cite{Gebresenbut} the ferrimagnetic distribution is realized. In the former, the total magnetization per the icosahedron is zero while in the latter, the total magnetization per the icosahedron is not zero. In these materials, the positive Curie-Weiss temperature was observed in the temperature dependence of the inverse of the magnetic susceptibility. This indicates the FM interaction works between the magnetic moments between the Tb (Ho) sites. 

Theoretical studies on magnetism of QCs and ACs have been stimulated by these experimental observations~\cite{Axe2001,Wessel2003,Kons2005,Jan2007,Hucht2011,Thiem2015,Komura2016,Sugimoto,Koga2017,Koga2020,Miyazaki2020,STS2021a,Koga2021,WK2021,WPNAS,WSR,Koga2022,Sugimoto2024,Ovarngard2024,WYTF2025}. To understand the magnetism in Gd- and Eu-based ACs where the effect of the crystalline electric field (CEF) is considered to be irrelevant, the model calculations were performed~\cite{Miyazaki2020,Sugimoto2024,Ovarngard2024}. 
Recently, Monte Carlo simulation of the ferromagnetic transition in icosahedral QC has been performed~\cite{WYTF2025}, which has attracted much attention. 

As for the rare-earth-based ACs except for Gd and Eu where the CEF is effective, 
to explain the measured magnetic structures as well as to predict new magnetic structures for future measurements, theoretical study is highly desired. 
As a first step of analysis of the strongly correlated electron state, magnetic anisotropy arising from the CEF at the rare-earth site is necessary to be clarified. 
However, the CEF in the rare-earth based icosahedral QCs and ACs has remained obscure because the well established CEF theory based on the crystallographic point group cannot be applied to the QC and AC owing to the five-fold symmetry around the rare-earth atom [see Fig.~\ref{fig:Tb_local}(f)]. 
Recently, the experimental analyses on the CEF in the rare-earth based 1/1 ACs have been reported on Zn$_{85.5}$Sc$_{11}$Tm$_{3.5}$~\cite{Jazbec}, Cd$_{6}$Tb~\cite{Das}, and Au$_{70}$Si$_{17}$Tb$_{13}$~\cite{Sato2019}. 

Theoretically, lack of the CEF theory under the five-fold rotation symmetry has prevented us from understanding the magnetism in the rare-earth based QCs and ACs, so far. 
A phenomenological model calculation for the magnetism was performed by assuming the magnetic anisotropy arising from the CEF at the rare-earth site in the 1/1 AC~\cite{Sugimoto}. 

Recently, the theory of the CEF in the QC and AC has been developed on the basis of the point charge model~\cite{WK2021}. 
This has made it possible to analyze the CEF in the rare-earth based QC and AC microscopically. 
Then, the CEF in the Au-SM-Tb QC and AC has been analyzed theoretically~\cite{WPNAS,WSR}. Interestingly, it has been shown that the magnetic easy axis for the CEF ground state at the Tb site is lying in the mirror plane, as illustrated in Fig~\ref{fig:Tb_local}(f). 
Namely, in the local coordinate, the $yz$ plane is the mirror plane, where  
the angle $\theta$ is defined between the magnetic easy axis ${\bm J}$ for the CEF ground state and the pseudo 5-fold axis ($z$ axis). 
It has been theoretically shown that as the ratio of the valences of the screened ligand ions $\alpha\equiv Z_{\rm SM}/Z_{\rm Au}$ increases, the magnetic easy axis, i.e., ${\bm J}$ in Fig~\ref{fig:Tb_local}(f),   rotates to the clockwise direction in the $yz$ plane and finally approaches the $z$ axis, i.e., the pseudo five-fold axis~\cite{WPNAS,WSR}. 
Here, $Z_{\rm SM}$ and $Z_{\rm Au}$ are the valences of the SM and Au ions respectively.

It is noted that the analysis of the CEF in the Tb-based 1/1 AC~\cite{WPNAS} was performed on the basis of the atomic coordinate of the 1/1 AC Au$_{70}$Si$_{17}$Tb$_{13}$~\cite{Hiroto}, where the existence ratio of the atoms at the cluster center shown in Fig.~\ref{fig:Tb_local}(a) is low (less than $23\%$~\cite{Hiroto}). Hence, the atoms at the cluster center were ignored as a first step of analysis in Ref.~\citen{WPNAS}, as shown in Fig.~\ref{fig:Tb_local}(f). 

By taking into account the uniaxial anisotropy arising from the CEF, the effective model was constructed, which will be explicitly explained in the following section II as Eq.~(\ref{eq:H}). Then, by numerical calculations, the ground state phase diagram of the effective model (\ref{eq:H}) on the icosahedron was determined for the FM interaction~\cite{WPNAS} and the antiferromagnetic (AF) interaction~\cite{WSR}. 

In refs.~\citen{WPNAS} and \citen{WSR}, the effective model was applied to the 1/1 AC and the ground states were also discussed. 
Namely, the magnetic ground state on single icosahedron was obtained by performing the numerically-exact calculation for the FM interaction~\cite{WPNAS} and the AFM interaction~\cite{WSR}. Then, by comparing which energy is lower, the uniform distribution of the magnetic states on the icosahedrons located at the center and corner of the bcc unit cell of the 1/1 AC, or the antiferromagnetic distribution, 
the ground state phase diagram was discussed 
in refs.~\citen{WPNAS} and \citen{WSR}. 

In this study, we determine the ground-state phase diagram of the effective model in the 1/1 AC with the uniaxial magnetic anisotropy arising from the CEF 
reported in Refs.~\citen{WPNAS} and \citen{WSR} 
by performing the 
numerically-exact calculation 
without assuming the magnetic state on the icosahedron. 
We show that eight kinds of noncollinear and noncoplanar magnetic structures are stabilized in the ground-state phase diagram for the FM interactions, for each of which we identify magnetic space group and degeneracy. 
The results explain the whirling-anti-whirling order observed in the 1/1 AC Au$_{72}$Al$_{14}$Tb$_{14}$~\cite{Sato2019} and the uniform non-collinear ferrimagnetic order observed in the 1/1 AC Au$_{70}$Si$_{17}$Tb$_{13}$~\cite{Hiroto}.
Moreover,  
we also find the other types of the non-collinear and non-coplanar magnetic structures. 
We clarify the property of each magnetic ground state by revealing not only the magnetic structure but also the topological character.

The organization of this paper is as follows.
In section II, we introduce the effective model for magnetism in the rare-earth based 1/1 AC. In section III, the results of the numerical calculation are presented. In section IV, the discussion
is given. In section V, we summarize the paper. 

\section{Effective model}

We consider the effective model taking into account the magnetic anisotropy arising from the CEF~\cite{WSR} as
\begin{eqnarray}
H=-\sum_{\langle i,j\rangle}J_{ij}\hat{\bm J}_i\cdot\hat{\bm J}_{j},
\label{eq:H}
\end{eqnarray}
where $\hat{\bm J}_i$ is the unit vector operator parallel or anti-parallel to the magnetic easy axis~\cite{Sato2019}. 
The interaction $J_{ij}$ is taken as the nearest-neighbor (N.N.) interaction $J_1$ and the next-nearest neighbor (N.N.N.) interaction $J_2$ not only for the intra icosahedron but also for the inter icosahedron.
In this study, we employ the lattice structure of the rare-earth atom in the 1/1 AC identified experimentally in Au$_{70}$Si$_{17}$Tb$_{13}$~\cite{Hiroto}.
The crystal structure is cubic (space group No. 204, $Im\bar{3}$, $T_h^{5}$). 

\begin{figure}[tb]
\includegraphics[width=7cm]{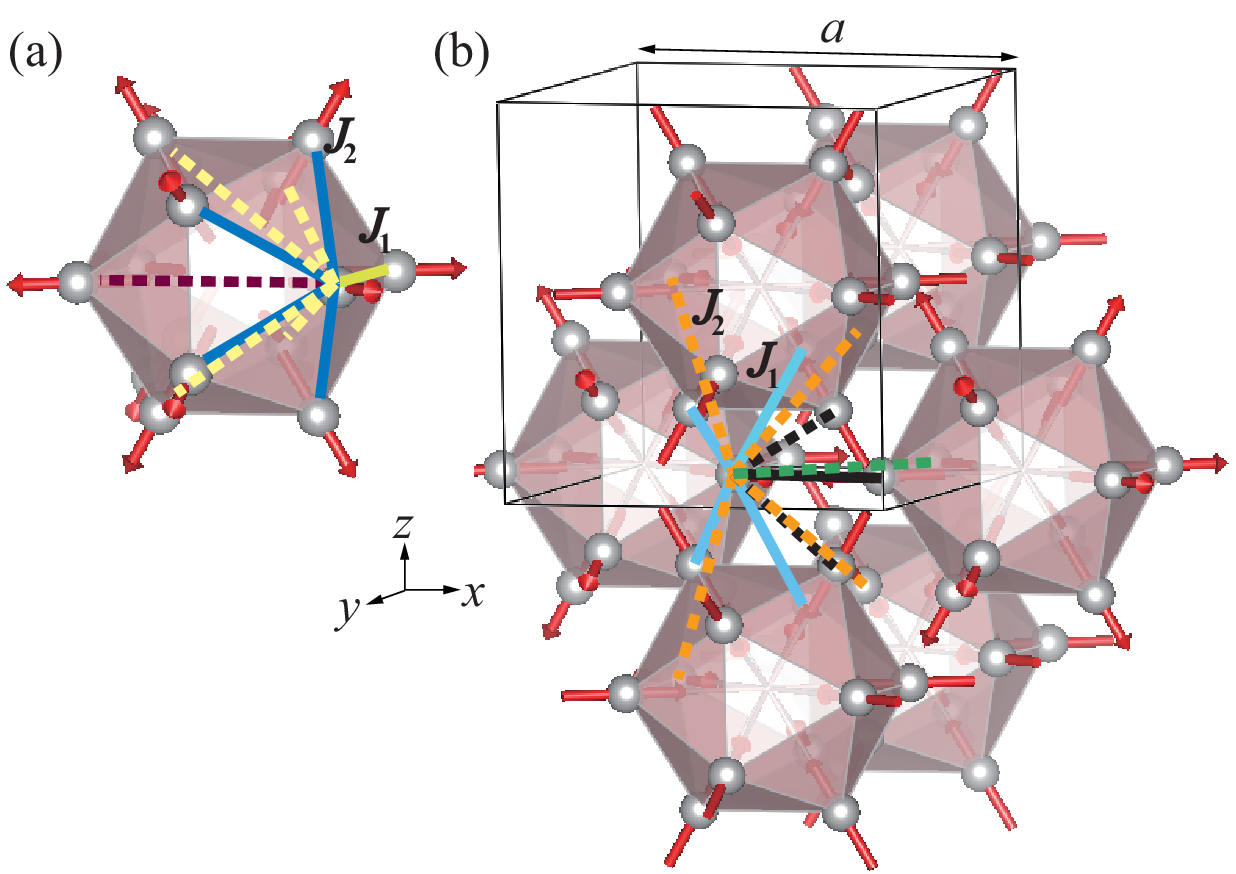}%
\caption{(Color online) 
The nearest-neighbor (N.N.) interactions $J_1$  (solid line) and next-nearest-neighbor (N.N.N.) interactions $J_2$  (dashed line) for (a) the intra icosahedron and  (b) the inter icosahedron are illustrated. 
In (b), the frame box represents the unit cell of the bcc lattice of the 1/1 AC with the lattice constant $a$. 
}
\label{fig:lattice}
\end{figure}

The interaction $J_{ij}$ set in this study is explained 
in Fig.~\ref{fig:lattice} where the bonds for $J_1$ are illustrated as bold solid lines and the bonds for $J_2$ are illustrated as bold dashed lines each color of which is noted in (  ) below.
As for the intra icosahedron interaction, 
$J_1$ is set for the 5 bonds for each Tb site with the bond length $0.374a$ (1 bond 
denoted by yellow line) and $0.378a$ (4 bonds 
denoted by blue lines) and $J_2$ is set for the 5 bonds for each Tb site with the bond length $0.610a$ (4 bonds 
denoted by yellow dashed lines) and $0.612a$ (1 bond 
denoted by purple dashed line). 
Here, $a$ is the lattice constant of the 1/1 AC [see Fig.~\ref{fig:lattice}(a)]. 
As for the inter icosahedron interaction, $J_1$ is set for the 5 bonds for each Tb site with the bond length $0.368a$ (4 bonds 
denoted by light blue lines) and $0.388a$ (1 bond 
denoted by light blue lines) and $J_2$ is set for the 7 bonds with the bond length $0.528a$ (2 bonds 
denoted by black dashed lines), $0.530a$ (4 bonds denoted by orange dashed lines), and $0.539a$ (1 bond 
denoted by green dashed line) [see Fig.~\ref{fig:lattice}(b)].

In this study, we analyze the case for the FM interaction in the model (\ref{eq:H}) on the 1/1 AC.
We consider the unit cell denoted as the frame box in Fig.~\ref{fig:lattice} with the two icosahedrons which contains 24 Tb sites under the periodic boundary condition.
We have performed the numerically-exact calculation for the model (\ref{eq:H}) to obtain the ground state. Namely, we have calculated the expectation values of the energy of the model (\ref{eq:H}) by all the eigenstates and have obtained the ground state with the lowest energy. 
It is noted that our method is equivalent to the exact diagonalization calculation to obtain numerically exact ground state.

We calculate the topological charge of the magnetic texture defined on the icosahedron which was introduced in ref.~\citen{WPNAS}. 
The topological charge is defined as 
the solid angle divided by the surface area of the unit sphere: 
\begin{eqnarray}
n=\frac{\Omega}{4\pi},
\label{eq:tp}
\end{eqnarray}
where $\Omega$ is the solid angle subtended by the 12 magnetic moments on the icosahedron 
\begin{eqnarray}
\Omega=\sum_{i,j,k\in {\rm IC}}\Omega_{ijk}.
\label{eq:SI_IC}
\end{eqnarray}
Here, $\Omega_{ijk}$ is the solid angle subtended by the three magnetic moments at each triangular surface of the icosahedron, $i$, $j$, and $k$ sites 
\begin{eqnarray}
\Omega_{ijk}=2\tan^{-1}\left[\frac{\chi_{ijk}}{1+{\bm J}_i\cdot{\bm J}_j+{\bm J}_j\cdot{\bm J}_k+{\bm J}_k\cdot{\bm J}_i}\right]
\label{eq:SI_tri}
\end{eqnarray}
\cite{Eriksson}, where $\chi_{ijk}$ is the scalar chirality defined by
\begin{eqnarray}
\chi_{ijk}={\bm J}_i\cdot({\bm J}_j\times{\bm J}_k). 
\label{eq:SC}
\end{eqnarray}
%


The total chirality is defined as
\begin{eqnarray}
{\bm \chi}^{\rm T}=\sum_{\langle i,j,k\rangle}\chi_{ijk}\hat{n}_{ijk},
\label{eq:chi_T}
\end{eqnarray}
where the summation is taken over all the three sites on the icosahedron and $\hat{n}_{ijk}$ denotes the surface normal~\cite{Aoyama2021}. 
The total chirality plays a role as the emergent fictious magnetic field, which gives rise to the topological Hall effect via the relation~\cite{Tatara}
\begin{eqnarray}
\sigma_{\mu\nu}^{\rm T}\propto\epsilon_{\mu\nu\rho}\chi_{\rho}^{\rm T},
\label{eq:sigma}
\end{eqnarray}
where $\epsilon_{\mu\nu\rho}$ is the Levi-Civita symbol
\begin{eqnarray}
\epsilon_{\mu\nu\rho}=
\left\{
\begin{array}{rl}
1, & (\mu,\nu,\rho)=(1,2,3), (2,3,1), (3,1,2) \\
-1, & (\mu,\nu,\rho)=(2,1,3), (1,3,2), (3,2,1) \\
0, & {\rm otherwise}. 
\end{array}
\right.
\label{eq:LC}
\end{eqnarray}

The total magnetic moment per the icosahedron is defined by
\begin{eqnarray}
{\bm J}_{\rm IC}=\sum_{i=1}^{12}{\bm J}_{i}. 
\label{eq:J_IC}
\end{eqnarray}

We have performed the numerical calculation by inputting various $J_2 (>0)$ and $J_1 (>0)$ with the anisotropy angle $\theta$. 
For the obtained ground state, we have calculated the topological charge $n$ in Eq.~(\ref{eq:tp}), total chirality ${\bm \chi}^{\rm T}$ in Eq.~(\ref{eq:chi_T}), and the total magnetic moment per the icosahedron ${\bm J}_{\rm IC}$ in Eq.~(\ref{eq:J_IC}).

\section{Results}

\subsection{Ground state phase diagram and ordering property at zero field}

\begin{figure}[tb]
\includegraphics[width=8cm]{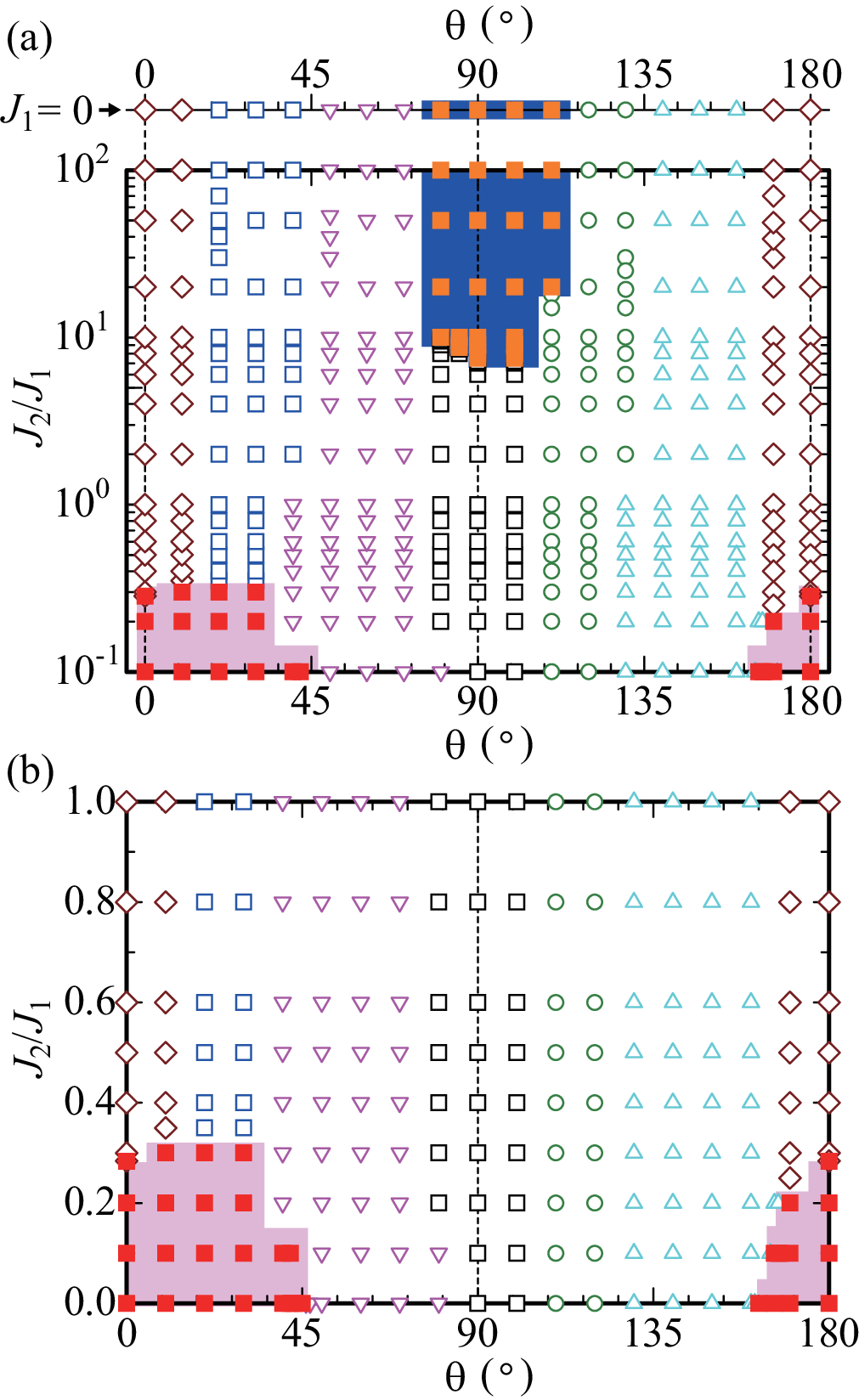}
\caption{(Color online) (a) 
The ground-state phase diagram in the 1/1 AC  (a) for $J_1=0$ (top) and in the plane of $J_2/J_1$ and the magnetic anisotropy angle $\theta~(^{\circ})$ (lower panel). 
The open symbol denotes the ferrimagnetic order where the magnetic states on the icosahedrons located at the center and the corner of the bcc unit cell are distributed uniformly. 
The filled symbol denotes the antiferromagnetic order where the magnetic states on the icosahedrons located at the center and the corner of the bcc unit cell are distributed in the antiferromagnetic alignment. 
The vertical dashed lines at $\theta=0^{\circ}$, $90^{\circ}$, and $180^{\circ}$ are guides for the eyes. 
(b) The enlargement for $0\le J_2/J_1 \le 1$.
In (a) and (b), the pink regions denote the ordered states with the topological charge $|n|=1$ on the icosahedron. In (a), the blue region denotes the ordered state with the topological charge $|n|=3$ on the icosahedron. 
}
\label{fig:PD}
\end{figure}

The results are summarized as the ground-state phase diagram in Fig.~\ref{fig:PD}. In Fig.~\ref{fig:PD}(a), the phase diagram for $J_1=0$ is shown at the top panel. In the lower panel, the phase diagram  in the plane of $J_2/J_1$ and the anisotropy angle $\theta$ is shown.  
The enlargement for $0\le J_2/J_1\le 1$ is presented in Fig.~\ref{fig:PD}(b).

\begin{figure}[h]
\includegraphics[width=7cm]{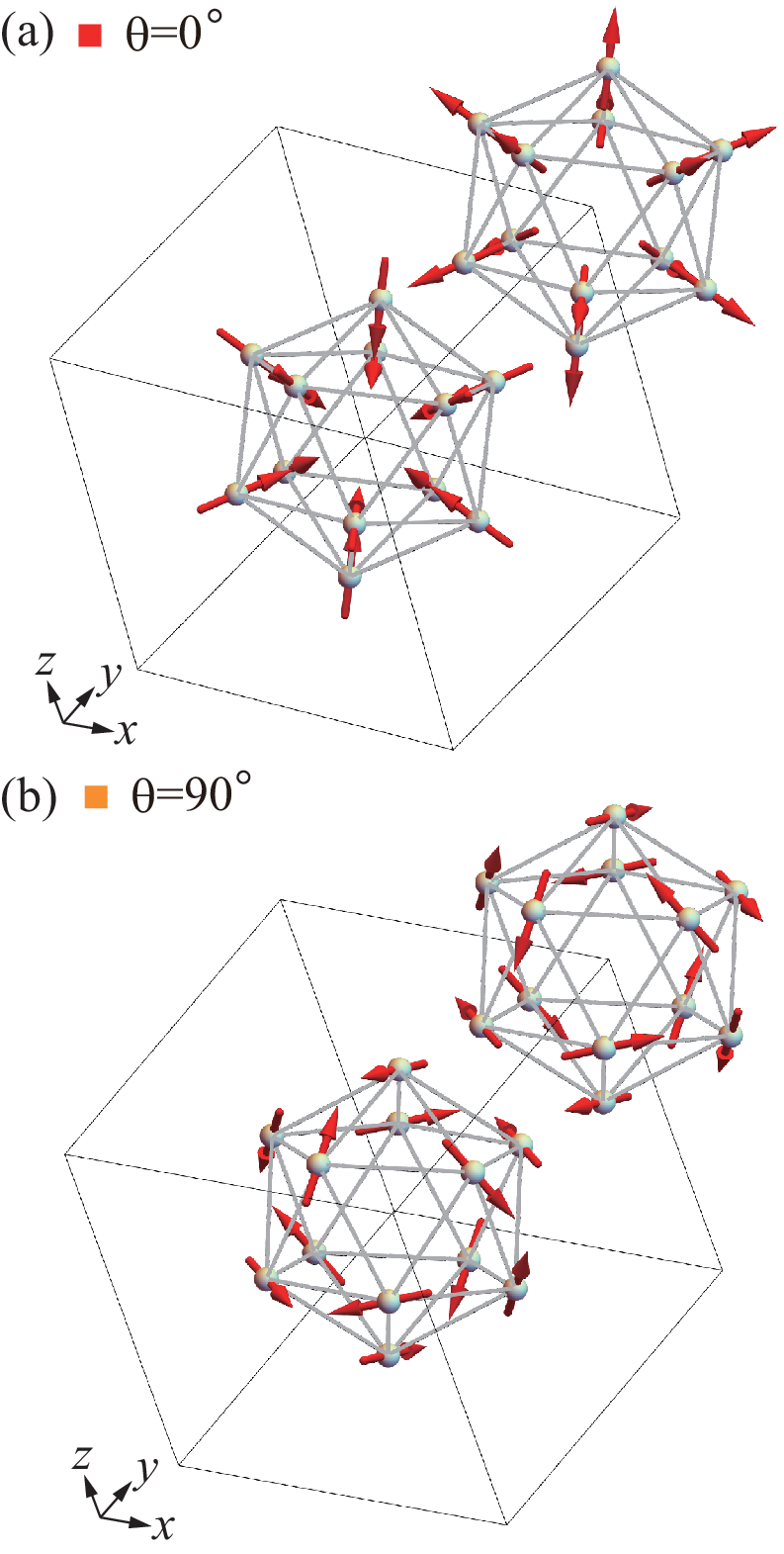}
\caption{(Color online) The antiferromagnetic orders on the icosahedrons located at the center and the corner of the bcc unit cell in the 1/1 AC. 
(a) The anti-hedgehog state at the center icosahedron and the hedgehog state at the corner icosahedron with the anisotropy angle $\theta=0^{\circ}$. 
(b) The anti-whirling state at the center icosahedron and the whirling state at the corner icosahedron with the anisotropy angle $\theta=90^{\circ}$. }
\label{fig:H_AH_W_AW}
\end{figure}

We obtained the eight kinds of non-collinear magnetic ground states in Fig.~\ref{fig:H_AH_W_AW}, whose magnetic space groups are identified as $I_{\rm P}m'\bar{3'}$, $C2'/m'$ and $R\bar{3}$. From a symmetry view point,  on the lattice characterized by the space group $Im\bar{3}$, the magnetic phases with the magnetic space groups $C2'/m'$ and $R\bar{3}$ can appear with the magnetic propagation vector $(000)$ while the magnetic phase with the magnetic space group $I_{\rm P}m'\bar{3'}$ can also appear with the magnetic propagation vector $(111)$. Then, the former and the latter phases are referred to as the ferrimagnetic phase and the antiferromagnetic phase, respectively.

\begin{figure*}[h]
\includegraphics[width=15cm]{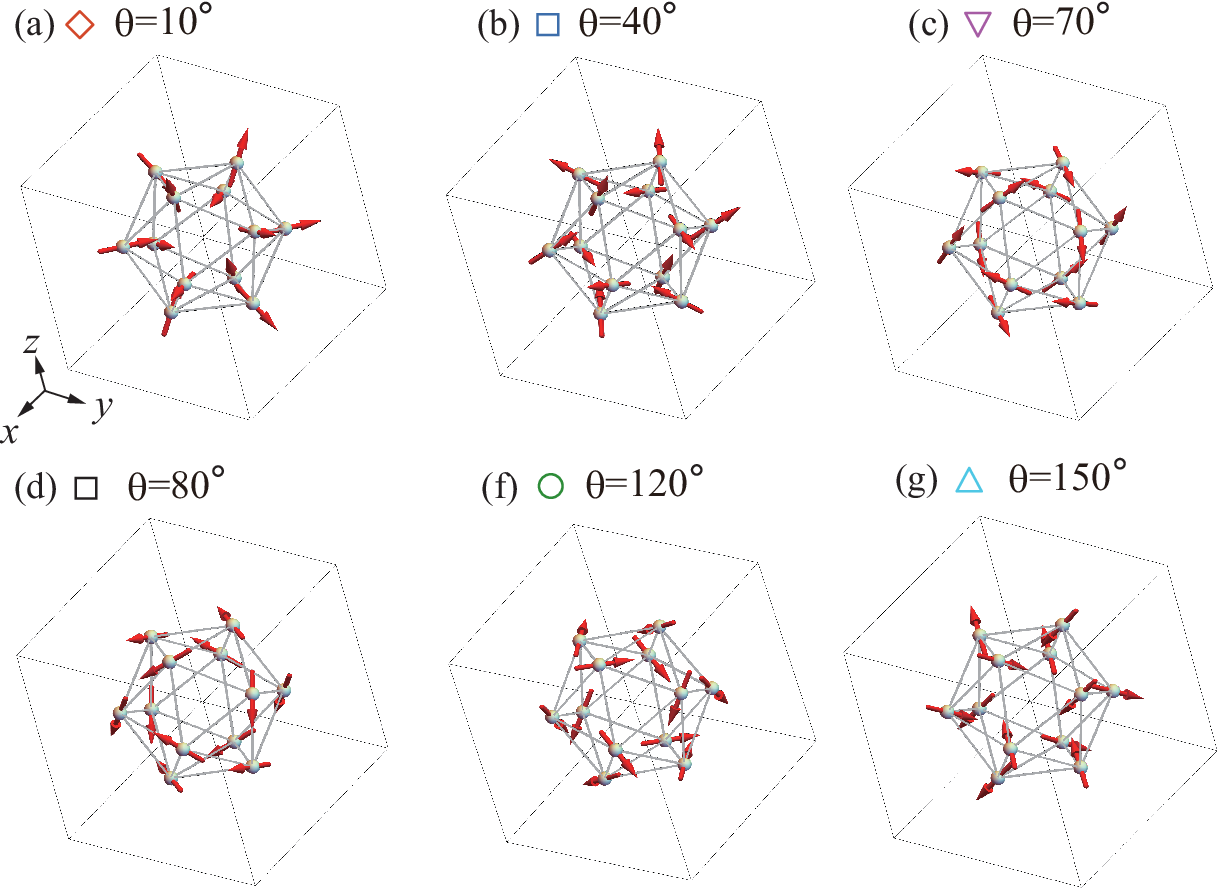}
\caption{(Color online) The ferrimagnetic orders on the icosahedrons located at the center and the corner of the bcc unit cell in the 1/1 AC.
The alignment of the magnetic moments on the central icosahedron is viewed from the (111) direction. 
}
\label{fig:uniform}
\end{figure*}

The filled symbols represent that the antiferromagnetic states on the icosahedrons located at the center and corner of the bcc unit cell of the 1/1 AC are stabilized as the ground state (see Fig.~\ref{fig:H_AH_W_AW}), while the open symbols represent the ferrimagnetic state stabilized as the ground state (see Fig.~\ref{fig:uniform}). 
Here, both figures are the views from the three-fold axis, i.e., the (111) direction. 
In Fig.~\ref{fig:PD}(a), the antiferromagnetic orders are realized in the regions for large $J_2/J_1$ near $\theta=90^{\circ}$ and for small $J_2/J_1$ near $\theta=0^{\circ}$ and $\theta=180^{\circ}$. 

In the previous study of the CEF in the Au-SM-Tb systems, it was shown theoretically that the parameter region $0\le\alpha\le 4$ corresponds to the magnetic anisotropy angle $15^{\circ}\lsim\theta\lsim 90^{\circ}$ at each Tb site~\cite{WSR}. Here, we determine the phase diagram for $0\le\theta\le 180^{\circ}$ to get insight into the general feature of the model (\ref{eq:H}). 
For each magnetic ground state, 
the magnetic-moment inverted state is energetically degenerate to the original state. Hence, each magnetic ground state for $\theta=0^{\circ}$ is the magnetic-moment inverted state of each ground state for $\theta=180^{\circ}$. Hence, the same symbols appear at $\theta=0^{\circ}$ and $180^{\circ}$ in Fig.~\ref{fig:PD}.

First, let us focus on the two limiting cases in the model Eq.~(\ref{eq:H}), i.e., the case with the N.N. interaction and the case with the N.N.N. interactions. The former phase diagram is presented as the basal line of Fig.~\ref{fig:PD}(b) where the antiferromagnetic orders are realized around $\theta=0^{\circ}$ and $180^{\circ}$ while the ferrimagnetic orders are realized in between. The latter phase diagram is displayed as the top line of Fig.~\ref{fig:PD}(a) where the antiferromagnetic order is realized around $\theta=90^{\circ}$ while the ferrimagnetic orders are realized around $\theta=0^{\circ}$ and $180^{\circ}$. Figure~\ref{fig:PD}(b) and the lower panel of Fig.~\ref{fig:PD}(a) display the $J_2/J_1$ variation of the phases stabilized between the two limits. In the following subsections, the property of each phase in the  antiferromagnetic orders and the ferrimagnetic orders will be explained. 

\subsubsection{Antiferromagnetic states}

In Fig.~\ref{fig:PD}, the filled symbols represent the ordered states where the antiferromagnetic distribution of the magnetic state on the icosahedron is realized at the center and corner of the unit cell. In this subsection, the property of each ordered state is discussed. 

The red-filled-square symbol in Fig.~\ref{fig:PD} represents the hedgehog-anti-hedgehog order as shown in Fig.~\ref{fig:H_AH_W_AW}(a). 
In the hedgehog state, all the magnetic moments at the 12 vertices of the icosahedron are directed outward from the icosahedron, while in the anti-hedgehog state, all the magnetic moments are directed inward from the icosahedron. 
For the magnetic anisotropy angle $\theta=0^{\circ}$ at each site, the hedgehog-anti-hedgehog order is realized for $J_2/J_1<0.284$. Namely, the strong FM N.N. interaction stabilizes this order. 
We confirmed that for $0\le\theta\le 45^{\circ}$ and for $162^{\circ}\le\theta\le 180^{\circ}$ the hedgehog-anti-hedgehog order is realized in the small-$J_2/J_1$ region [see Fig.~\ref{fig:PD}(b)]. 

As seen in Fig.~\ref{fig:H_AH_W_AW}(a), the 3-fold symmetry holds along the (111) direction with the inversion center with respect to the atomic positions with opposite magnetic-moment directions. The four 3 fold axes cross at the inversion center which is located at the center of the bcc unit cell, which implies the cubic-lattice symmetry. Moreover, the mirror symmetry with time reversal operation holds with respect to the mirror plane perpendicular to the 2-fold axis. The magnetic state at the corner icosahedron is obtained by the $(1/2,1/2,1/2)$ translation of the magnetic state at the center icosahedron in the bcc unit cell with time reversal operation. These symmetry operations conclude that the magnetic space group of the hedgehog-anti-hedgehog state is $I_{\rm P}m'\overline{3'}$ (No. 204.5.1534 in the OG notation and No. 201.21 $PIn\bar{3}$ in the BNS notation)~\cite{OG,BNS}. 

The total magnetic moment per the icosahedron is zero ${\bm J}_{\rm IC}={\bf 0}$ for the hedgehog state and the anti-hedgehog state. 
The total chirality ${\bm \chi}^{\rm T}={\bf 0}$ for the hedgehog state and also for the anti-hedgehog state. 
The magnetic-moment inverted state in Fig.~\ref{fig:H_AH_W_AW}(a), i.e., anti-hedgehog-hedgehog state, is energetically degenerate. 
The hedgehog state on the icosahedron located at the corner of the bcc unit cell is characterized by the topological charge $n=+1$ and the anti-hedgehog state on the icosahedron located at the center of the unit cell is characterized by the topological charge $n=-1$. 
The magnetic ground state on the icosahedron with the topological charge $|n|=1$ is shown as the filled-pink region in Figs.~\ref{fig:PD}(a) and \ref{fig:PD}(b). 
These characteristics are summarized in Table~\ref{tb:mag_state}.
It is noted that near the phase boundary the degeneracy increases because of the competition with the other ordered states. For example, at $\theta=45^{\circ}$ for $J_2/J_1=0$, the degeneracy of the ground states is 66 where the whiling-anti-whirling ordered states with degeneracy 2 are included. 

\begin{table*}
\caption{
Characteristics of each magnetic order realized in the 1/1 AC. 
The symbol in the 1st column denotes 
the magnetic state shown in Fig.~\ref{fig:H_AH_W_AW} or Fig.~\ref{fig:uniform}. 
The magnetic propagation vector ${\bm q}=\frac{2\pi}{a}(hkl)$, the magnetic space group in the OG notation, the independent magnetic position, degeneracy, the total magnetic moment vector per icosahedron ${\bm J}_{\rm IC}$, the total chirality $\chi^{\rm T}$, and topological charge 
On the 6th and 7th columns, the states with all the components of ${\bm J}_{\rm IC}$ and ${\bm \chi}^{\rm T}$ having opposite sign are energetically degenerate. 
The IC1 (IC2) is the icosahedron located at the corner (center) of the unit cell in Fig.~\ref{fig:H_AH_W_AW} and Fig.~\ref{fig:uniform}. 
The lattice constant $a$ is $a=14.725~{\rm \AA}$ for cubic unit cell (i.e., the side length of a cube in Fig.~\ref{fig:H_AH_W_AW} and Fig.~\ref{fig:uniform})~\cite{Hiroto}. 
}
\label{tb:mag_state}
\begin{tabular}{lccllcllr}
\hline
\hline
states &
$(hkl)$ & mag. & position  &  
\textrm{deg.}& \textrm{${\bm J}_{\rm IC}$} & \textrm{${\bm \chi}^{\rm T}$} &
\multicolumn{2}{c}{$n$}\\
& & sp. gr. & & & & & 
\multicolumn{1}{c}{\textrm{IC1}}&
\textrm{IC2}\\
\hline
red square
\\ Fig.~\ref{fig:H_AH_W_AW}(a)
 & (111) & $I_{\rm P}m'\bar{3}'$ & $(0.187,0.306,0)$ & 2 & {\bf 0} & {\bf 0} & 1 & -1\\
\hline
brown  & $(000)$ & $C2'/m'$ & $(0.187,0.306,0.187)$ & 12 & $(2.99, 5.65, 0.00)$ & $(2.72, 5.93, 0.00)^{\rm a}$ & 0 & 0 \\
 diamond & & & $(0,0.187,0.306)$ & & $(0.00, 2.99, 5.65)$ & $(0.00, 2.72, 5.93)$ & 0 & 0 \\
 Fig.~\ref{fig:uniform}(a) & & & $(0.306,0,0.119)$ & & $(-2.99, 5.65, 0.00)$ & $(-2.72, 5.93, 0.00)$ & 0 & 0 \\ 
& & &  $(0.306,0,0.493)$ & & $(0.00, 2.99,-5.65)$ & $( 0.00, 2.72,-5.93)$ & 0 & 0 \\ 
& & & & & $(5.65, 0.00, 2.99)$ & $( 5.93, 0.00, 2.72)$ & 0 & 0 \\ 
& & & & & $(5.65, 0.00,-2.99)$ & $( 5.93, 0.00,-2.72)$ & 0 & 0 \\ 
\hline 
blue 
  & $(000)$ & $C2'/m'$ & $(0.187,0.306,0.187)$ &12 & $(0.00, 5.05, 3.80)$ & $(0.00, 0.00, 6.35)^{\rm b}$ & 0 & 0 \\
square & & & $(0,0.187,0.306)$ & & $(0.00, 5.05,-3.80)$ & $(0.00, 0.00,-6.35)$ & 0 & 0 \\
Fig.~\ref{fig:uniform}(b)& & & $(0.306,0,0.119)$ & & $(5.05, 3.80, 0.00)$ & $(0.00, 6.35, 0.00)$ & 0 & 0 \\ 
& & & $(0.306,0,0.493)$ & & $(3.80, 0.00, 5.05)$ & $( 6.35, 0.00, 0.00)$ & 0 & 0 \\
& & &  & & $(5.05,-3.80, 0.00)$ & $(0.00,-6.35, 0.00)$ & 0 & 0 \\
& & &  & & $(3.80, 0.00,-5.05)$ & $(6.35, 0.00, 0.00)$ & 0 & 0 \\ 
\hline 
pink  
 & $(000)$ & $R\bar{3}$ & $(-0.170,0.103,0.265)$ & 8 & $(3.92, 3.92, 3.92)$ & $(-0.32,-0.32,-0.32)^{\rm c}$ & 0 & 0 \\
inverted & & & $(-0.427,-0.265,-0.162)$ & & $(3.92, 3.92,-3.92)$ & $(-0.32,-0.32, 0.32)$ & 0 & 0 \\
 triangle & & &  & & $(3.92,-3.92, 3.92)$ & $(-0.32, 0.32,-0.32)$ & 0 & 0 \\ 
 Fig.~\ref{fig:uniform}(c) & & &  & & $(-3.92, 3.92, 3.92)$ & $(0.32,-0.32,-0.32)$ & 0 & 0 \\ 
\hline 
black    
 & $(000)$ & $C2'/m'$ & $(0.187,0.306,0.187)$ & 12 & $(0.00,-5.51, 3.40)$ & $( 0.00,-2.04,-7.03)^{\rm d}$ & 0 & 0 \\
square & &  & $(0,0.187,0.306)$ & & $(0.00,-5.51,-3.40)$ & $( 0.00,-2.04, 7.03)$ & 0 & 0 \\
Fig.~\ref{fig:uniform}(d)& &  & $(0.306,0,0.119)$ & & $(3.40, 0.00,-5.51)$ & $(-7.03, 0.00,-2.04)$ & 0 & 0 \\
& & &  $(0.306,0,0.493)$ & &  $(5.51, 3.40, 0.00)$ & $(2.04,-7.03, 0.00)$ & 0 & 0 \\
& & & & & $(3.40, 0.00, 5.51)$ &  $(-7.03, 0.00, 2.04)$ & 0 & 0 \\
& & & & & $(5.51,-3.40, 0.00)$ &  $(2.04, 7.03, 0.00)$ & 0 & 0 \\
\hline
green    
 & $(000)$ & $C2'/m'$ & $(0.187,0.306,0.187)$ & 12 & $(0.00,-3.52,-5.42)$ & $( 0.00, 1.67, 0.00)^{\rm e}$ & 0 & 0 \\
circle & & & $(0,0.137,0.306)$ & & $(-3.52,-5.42, 0.00)$ & $( 1.67, 0.00, 0.00)$ & 0 & 0 \\
Fig.~\ref{fig:uniform}(e)& & & $(0306,0,0.119)$ & &$(0.00,-3.52, 5.42)$ & $( 0.00, 1.67, 0.00)$ & 0 & 0 \\
& & & $(0.306,0,0.493)$ & & $(3.52,-5.42, 0.00)$ & $(-1.67, 0.00, 0.00)$ & 0 & 0 \\
& & & & & $(5.42, 0.00,-3.52)$ & $( 0.00, 0.00, 1.67)$ & 0 & 0 \\
& & & & & $(5.42, 0.00, 3.52)$ & $( 0.00, 0.00,-1.67)$ & 0 & 0 \\
\hline
light blue  
 & $(000)$ & $R\bar{3}$ & $(-0.170,0.103,0.265)$ & 8 & $(-4.00,-4.00,-4.00)$ & $(-0.24,-0.24,-0.24)^{\rm f}$ & 0 & 0 \\
 triangle & & & $(-0.427,-0.265,-0.162)$ & & $( 4.00,-4.00,-4.00)$ & $( 0.24,-0.24,-0.24)$ & 0 & 0 \\
Fig.~\ref{fig:uniform}(f) & & & & & $(-4.00,-4.00, 4.00)$ & $(-0.24,-0.24, 0.24)$ & 0 & 0 \\
 & & & & & $(-4.00, 4.00,-4.00)$ & $(-0.24, 0.24,-0.24)$ & 0 & 0 \\
\hline
orange square 
\\
Fig.~\ref{fig:H_AH_W_AW}(b) & $(111)$ &  $I_{\rm P}m'\bar{3}'$ & $(0.187,0.306,0)$  & 2 & {\bf 0} & {\bf 0} & 3 & -3 
\\
\hline
\hline
\end{tabular}
\\
$^{\rm a}$ \footnotesize{$\theta=10^{\circ}$ case. For the $\theta$ dependence, see Fig.~\ref{fig:chi_T_bd}.}
\\
$^{\rm b}$ \footnotesize{$\theta=40^{\circ}$ case. For the $\theta$ dependence, see Fig.~\ref{fig:chi_T_bs}.}
\\
$^{\rm c}$ \footnotesize{$\theta=70^{\circ}$ case. For the $\theta$ dependence, see Fig.~\ref{fig:chi_T_x}.}
\\
$^{\rm d}$ \footnotesize{$\theta=90^{\circ}$ case. For the $\theta$ dependence, see Fig.~\ref{fig:chi_T_black_square}.}
\\
$^{\rm e}$ \footnotesize{$\theta=120^{\circ}$ case. For the $\theta$ dependence, see Fig.~\ref{fig:chi_T_gc}.}
\\
$^{\rm f}$ \footnotesize{$\theta=150^{\circ}$ case. For the $\theta$ dependence, see Fig.~\ref{fig:chi_T_lbt}.}
\end{table*}

The orange-filled-square symbol in Fig.~\ref{fig:PD} represents the whirling-anti-whirling order as shown in Fig.~\ref{fig:H_AH_W_AW}(b). 
As seen from the 3-fold axis e.g., the (111) direction in Fig.~\ref{fig:H_AH_W_AW}(b), the whirling magnetic moments are formed, which was actually identified by the neutron measurement in the 1/1 AC Au$_{72}$Al$_{14}$Tb$_{14}$~\cite{Sato2019}. 
In the vicinity of $\theta=90^{\circ}$ for the large N.N.N. $J_2/J_1$ region in Fig.~\ref{fig:PD}, the whirling-anti-whirling order is stabilized in the 1/1 AC. 
The total magnetic moment per the icosahedron is zero ${\bm J}_{\rm IC}={\bf 0}$ for the whirling-moment state and the anti-whirling-moment state. 
The total chirality is zero ${\bm \chi}^{\rm T}={\bf 0}$ for the whirling-moment state and also for the anti-whirling-moment state. 
The whirling-moment state at the corner of the bcc unit cell is characterized by the topological charge $n=+3$ and anti-whirling moment state at the center of the bcc unit cell is characterized by $n=-3$~\cite{WPNAS}. 
The magnetic ground state on the icosahedron with the topological charge $|n|=3$ is shown as the blue region in Fig.~\ref{fig:PD}(a).
The magnetic-moment inverted state in Fig.~\ref{fig:H_AH_W_AW}(b), i.e., the ordered state of the anti-whirling-moment state at the corner icosahedron and the whirling-moment state at the center icosahedron is energetically degenerate (see Table~\ref{tb:mag_state}). 

The magnetic space group of the whiling-anti-whirling state is $I_{\rm P}m'\overline{3'}$ (No. 204.5.1534 in the OG notation and No. 201.21 $PIn\bar{3}$ in the BNS notation)~\cite{Labib}. The magnetic space group of the hedgehog-anti-hedgehog state shown in Fig.~\ref{fig:H_AH_W_AW}(a) and the whiling-anti-whirling state shown in Fig.~\ref{fig:H_AH_W_AW}(b) is the same.


\subsubsection{Ferrimagnetic states}

In Fig.~\ref{fig:PD}, the open symbols represent the ordered states where the magnetic state on the icosahedron is uniformly distributed at the center and corner of the unit cell. In this subsection, the property of each ordered state is discussed. 

\begin{figure}[h]
\includegraphics[width=6cm]{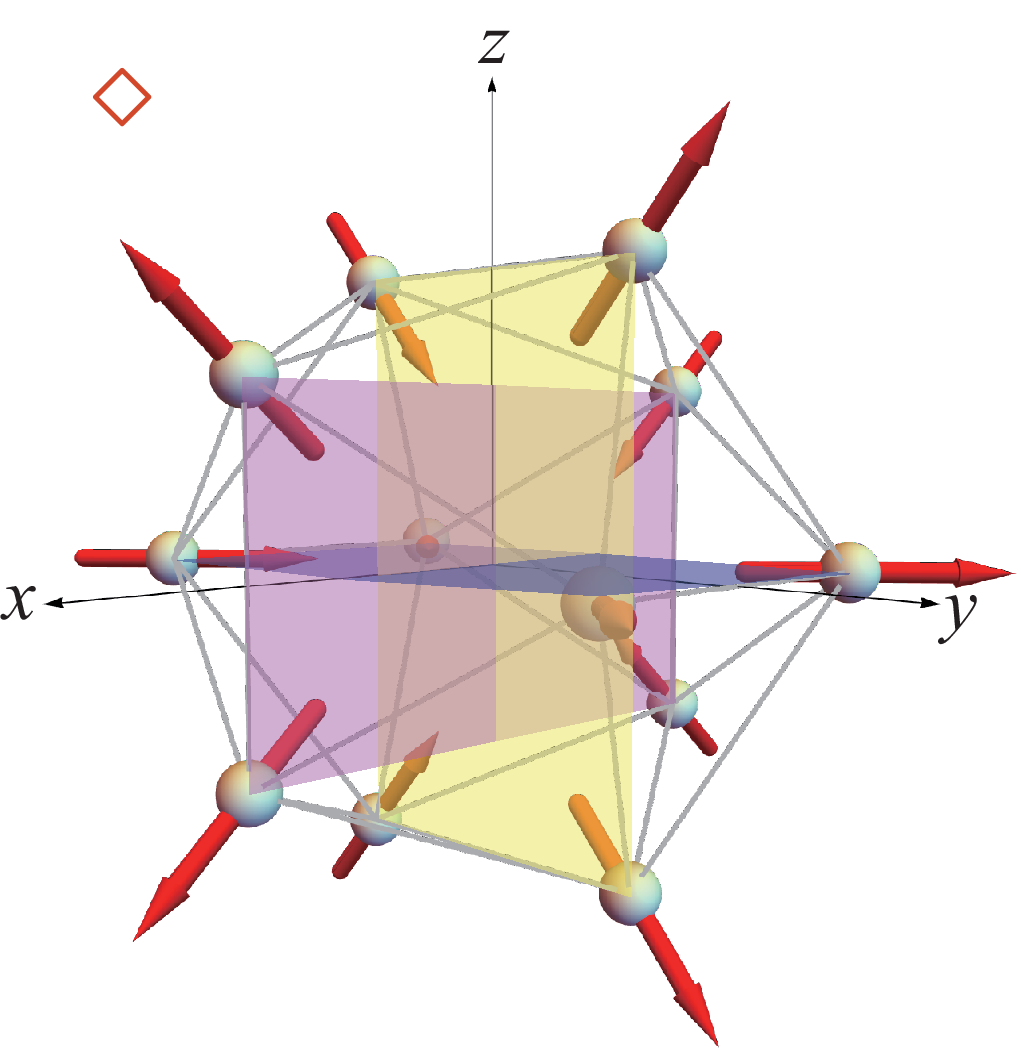}
\caption{(Color online) The non-collinear ferrimagnetic state for $\theta=10^{\circ}$ on the icosahedron denoted by the brown-diamond symbol in Fig.~\ref{fig:uniform}(a).}
\label{fig:brown_diamond_IC}
\end{figure}

The brown-diamond symbol in Fig.~\ref{fig:PD} represents the ferrimagnetic order of the non-collinear ferrimagnetic states shown in Fig.~\ref{fig:uniform}(a). 
Figure~\ref{fig:brown_diamond_IC} illustrates the magnetic structure on the icosahedron for $\theta=10^{\circ}$. 
This state can be obtained by inverting the 6 magnetic moments of the hedgehog state for $\theta=10^{\circ}$. 

The magnetic space group of this state is $C2'/m'$ (No. 12.5.70 in the OG notation and No. 12.62  $C2'/m'$ in the BNS notation) [50,51]. Namely, the alignment of the magnetic moments of this state is invariant under the 180$^{\circ}$ rotation around the $z$ axis with inversion of the magnetic moment. This state is also invariant under the mirror operation with respect to the mirror plane perpendicular to the $z$ axis. 

From the space group $Im\bar{3}$, one axis is singled out as a unique axis in monoclinic $C2'/m'$ and it turns out that the resulting magnetic moment is zero along this axis. The magnetic state shown in Fig.~\ref{fig:brown_diamond_IC} has the net magnetic moment of the icosahedron ${\bm J}_{\rm IC}=(2.99,5.65,0.00)$ and the total chirality ${\bm \chi}^{\rm T}=(2.72,5.93,0.00)$. 
From $Im\bar{3}$ to the monoclinic $C2'/m'$, the way of choosing the unique axis among $\pm{x}$, $\pm{y}$, $\pm{z}$ directions is six, which gives the six domain structures. Our exact numerical calculation gives the twelve degeneracy of the ground state, which corresponds to the above six degree with time reversal one (see Table~\ref{tb:mag_state}). On the 4th column of Table~\ref{tb:mag_state}, we list the independent magnetic position in unit of the lattice constant of the unit cell. On the 6th and 7th columns of Table~\ref{tb:mag_state}, we also list ${\bm J}_{\rm IC}$ and ${\bm \chi}^{\rm T}$ for each degenerate state, respectively. The first line of ${\bm J}_{\rm IC}$ and ${\bm \chi}^{\rm T}$ is the result of the magnetic structure shown in Fig.~\ref{fig:brown_diamond_IC} [same for other states shown in Figs.~\ref{fig:blue_sq_IC}, \ref{fig:ferri_IC}(a), \ref{fig:black_sq_IC}, \ref{fig:green_circle_IC}, and \ref{fig:light_blue_tri_IC}(a) below]. 



\begin{figure}[h]
\includegraphics[width=8cm]{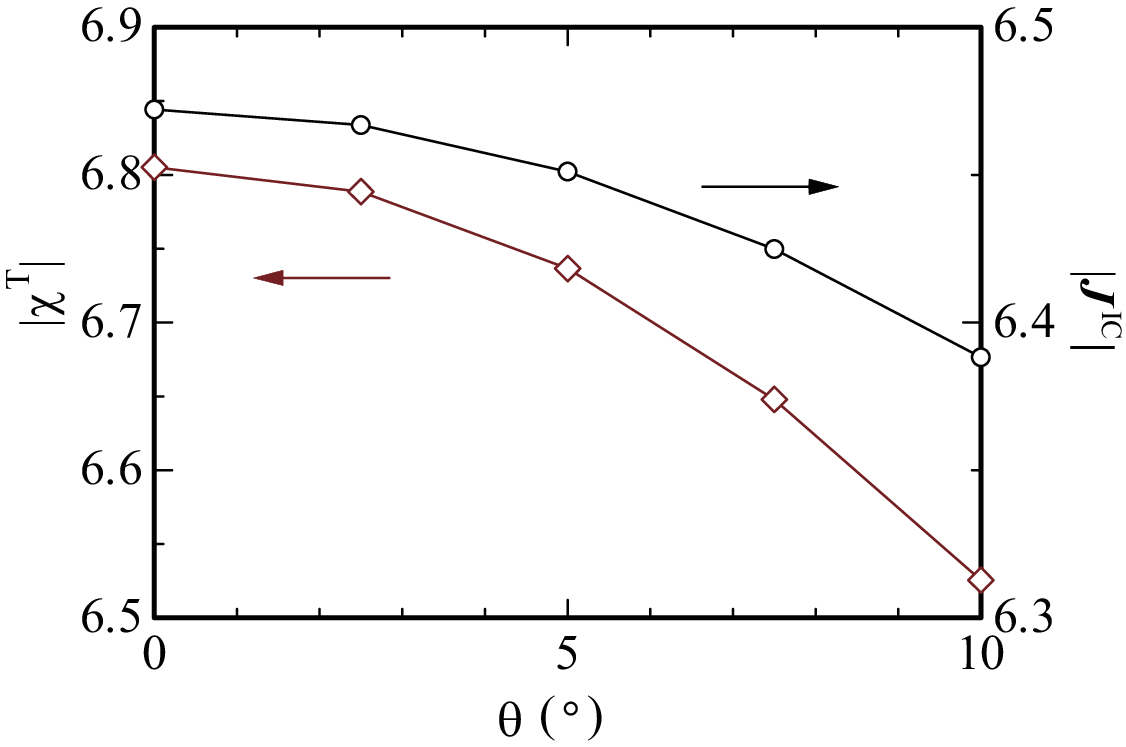}
\caption{(Color online) The anisotropy-angle $\theta$ dependence of $|{\bm \chi}^{\rm T}|$ (left axis) and $|{\bm J}_{\rm IC}|$ (right axis) in the magnetic state shown in Fig.~\ref{fig:uniform}(a). 
}
\label{fig:chi_T_bd}
\end{figure}

The anisotropy-angle $\theta$ dependences of the absolute values of $|{\bm \chi}^{\rm T}|$ and the total magnetic moment of the icosahedron $|{\bm J}_{\rm IC}|$ are shown in Fig.~\ref{fig:chi_T_bd}. 
As $\theta$ increases, $|{\bm \chi}^{\rm T}|$ and $|{\bm J}_{\rm IC}|$ decreases monotonically. 
The topological charge of this state is zero $n=0$. 
These characteristics are summarized in Table~\ref{tb:mag_state}.

The blue-square symbol shown in Fig.~\ref{fig:PD} represents the ferrimagnetic order of the non-collinear ferrimagnetic state shown in Fig.~\ref{fig:uniform}(b). 
Figure~\ref{fig:blue_sq_IC} shows the magnetic structure on the icosahedron for $\theta=40^{\circ}$. 

The magnetic space group of this state is $C2'/m'$ (No. 12.5.70 in the OG notation and No. 12.62  $C2'/m'$ in the BNS notation) [50,51]. Namely, the alignment of the magnetic moments of this state is invariant under the 180$^{\circ}$ rotation around the $x$ axis with inversion of the magnetic moment. This state is also invariant under the mirror operation with respect to the mirror plane perpendicular to the $x$ axis.

\begin{figure}[h]
\includegraphics[width=6cm]{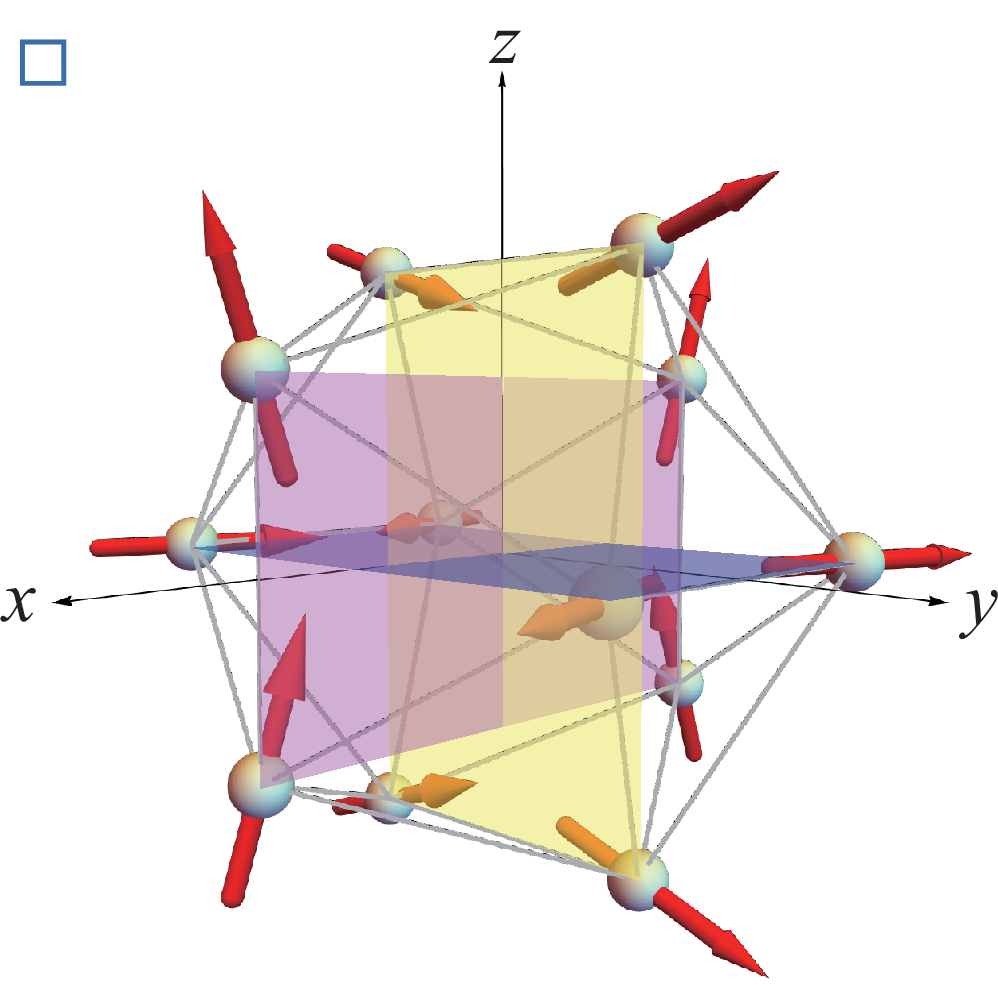}
\caption{(Color online) The non-collinear ferrimagnetic state for $\theta=40^{\circ}$ on the icosahedron denoted by the blue-square symbol in Fig.~\ref{fig:uniform}(b).}
\label{fig:blue_sq_IC}
\end{figure}

The magnetic state shown in Fig.~\ref{fig:blue_sq_IC} 
 has the net magnetic moment of the icosahedron ${\bm J}_{\rm IC}=(0.00,5.05,3.80)$ and the total chirality ${\bm \chi}^{\rm T}=(0.00,0.00,6.35)$.
 The characteristics of this state are listed in Table~\ref{tb:mag_state}.
  
\begin{figure}[h]
\includegraphics[width=8cm]{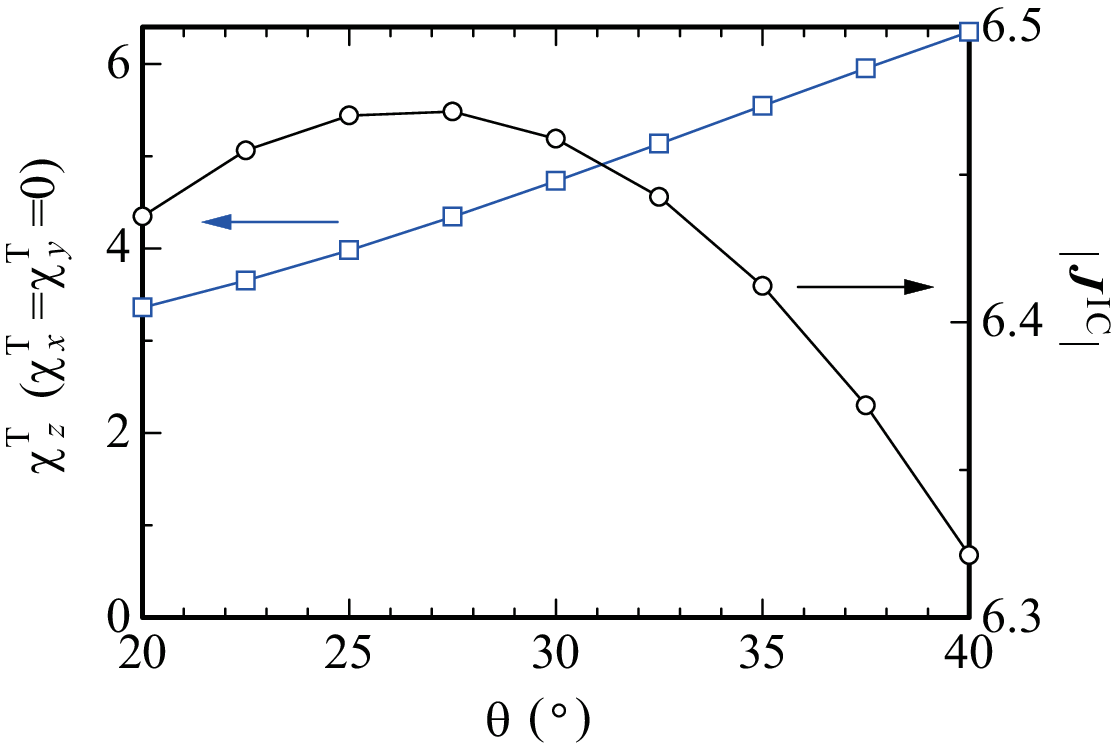}
\caption{(Color online) The anisotropy-angle $\theta$ dependence of $\chi^{\rm T}_{z}$ (left axis) and $|{\bm J}_{\rm IC}|$ (right axis) in the magnetic state shown in Fig.~\ref{fig:uniform}(b).}
\label{fig:chi_T_bs}
\end{figure}

\begin{figure}[h]
\includegraphics[width=8cm]{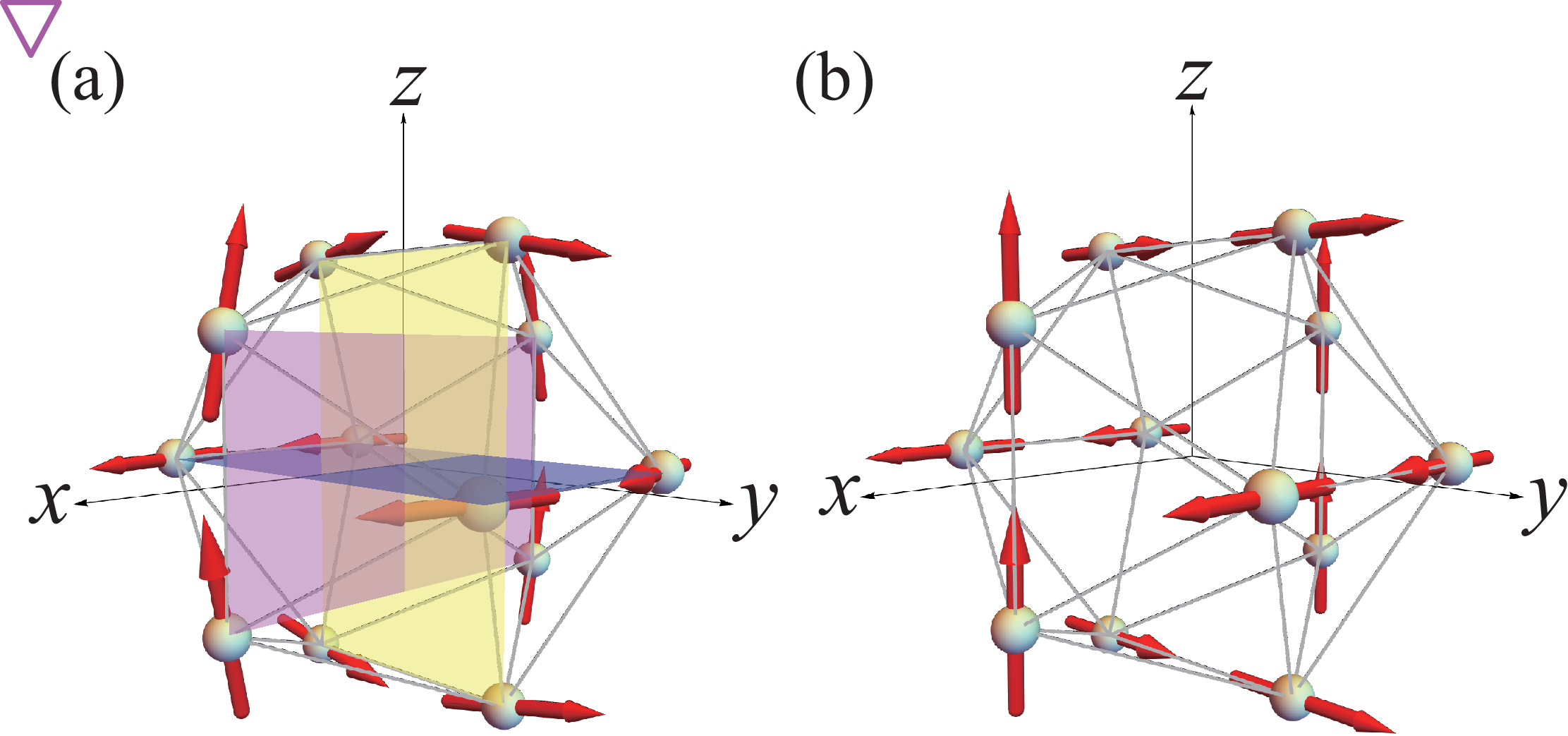}
\caption{(Color online) The non-collinear ferrimagnetic states for (a) $\theta=70^{\circ}$ (b) $\theta=58.28254^{\circ}$ on the icosahedron denoted by the pink-inverted-triangle symbol in Fig.~\ref{fig:uniform}(c).}
\label{fig:ferri_IC}
\end{figure}

The anisotropy-angle dependence of $\chi^{\rm T}_{z}$ and the absolute value of the total magnetic moment of the icosahedron $|{\bm J}_{\rm IC}|$ is shown in Fig.~\ref{fig:chi_T_bs}. 
As $\theta$ increases, $\chi^{\rm T}_{z}$ increases monotonically while the magnitude of the total magnetic moment has a maximum around $\theta=27^{\circ}$. 
The topological charge of this state is zero $n=0$. 



The pink-inverted-triangle symbol in Fig.~\ref{fig:PD} represents the ferrimagnetic order of the non-collinear ferrimagnetic state shown in Fig.~\ref{fig:uniform}(c).
Figure~\ref{fig:ferri_IC}(a) illustrates the magnetic structure on the icosahedron for $\theta=70^{\circ}$. 

The magnetic space group of this state is $R\bar{3}$ (No. 148.1.1247 in OG notation and 148.17 $R\bar{3}$ in BNS notation) [50,51]. Namely, the alignment of the magnetic moments of this state satisfies the 3-fold rotation symmetry along the (111) direction, which is invariant under inversion, as seen in Fig.5(c). From the space group $Im\bar{3}$, one 3-fold axis is singled out as a unique axis in rhombohedral  $R\bar{3}$ among four diagonal directions of the cubic ant it turns out that the resultant magnetic moment vector per icosahedron is along the unique 3-fold axis. For instance, the unique axis of the magnetic state shown in Fig.10(a) is the (111) direction, where the net magnetization per icosahedron is ${\bm J}_{\rm IC}=(3.92, 3.92, 3.92)$. The way of choosing the one 3-fold axis for rhombohedral lattice from cubic $Im\bar{3}$ is 4, i.e., four diagonal directions of cubic. This four degree and time reversal one yield $4\times 2=8$ gives eight types of domain structures. Our exact numerical calculation showed the eight degeneracy of the ground state, which confirms this.

This ferrimagnetic order appears in the region for $40^{\circ}\le\theta\le 80^{\circ}$ of the ground state phase diagram shown in Figs.~\ref{fig:PD}(a) and \ref{fig:PD}(b). 
The rectangles colored in pink, yellow, and purple in Fig.~\ref{fig:ferri_IC}(a) includes the mirror plane located at each vertex whose local coordinate is shown in Fig.~\ref{fig:Tb_local}(f). 
The total magnetic-moment vector at the four vertices of the purple, yellow, and pink rectangles are directed to the $x, y,$ and $z$ directions, respectively~\cite{Hiroto}. 
Namely, this state has the net magnetic moment along the (111) direction per the icosahedron. 


It is also noted that at $J_2/J_1=0$, we confirmed that the degeneracy is 64 for $50^{\circ}\le\theta\le80^{\circ}$ where the magnetic states with the degeneracy 8 listed in Table~\ref{tb:mag_state} are included.

\begin{figure}[h]
\includegraphics[width=8cm]{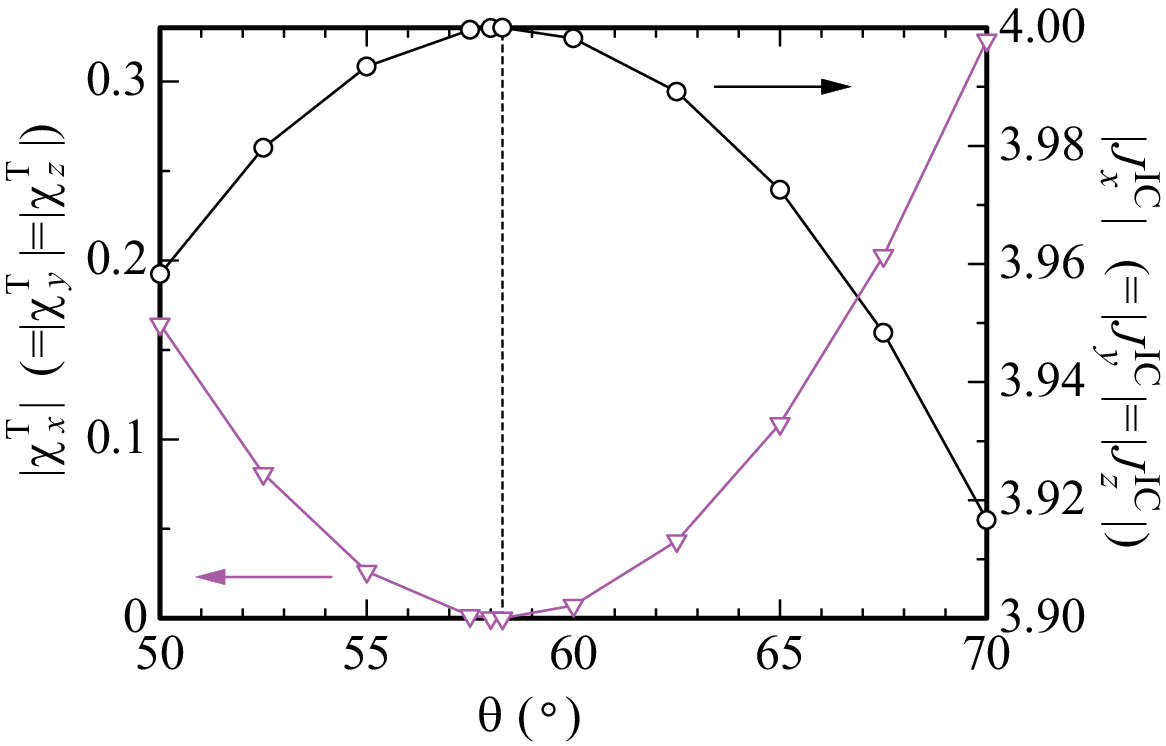}
\caption{(Color online) The anisotropy-angle $\theta$ dependence of $|\chi^{\rm T}_{\mu}|~(\mu=x, y, z)$ (left axis) and the component of $|J_{{\rm IC},\mu}|~(\mu=x, y, z)$ (right axis) in the non-collinear ferrimagnetic state. 
The vertical dashed line indicates $\theta=\theta_0$ (see text).}
\label{fig:chi_T_x}
\end{figure}

\begin{figure}
\includegraphics[width=6cm]{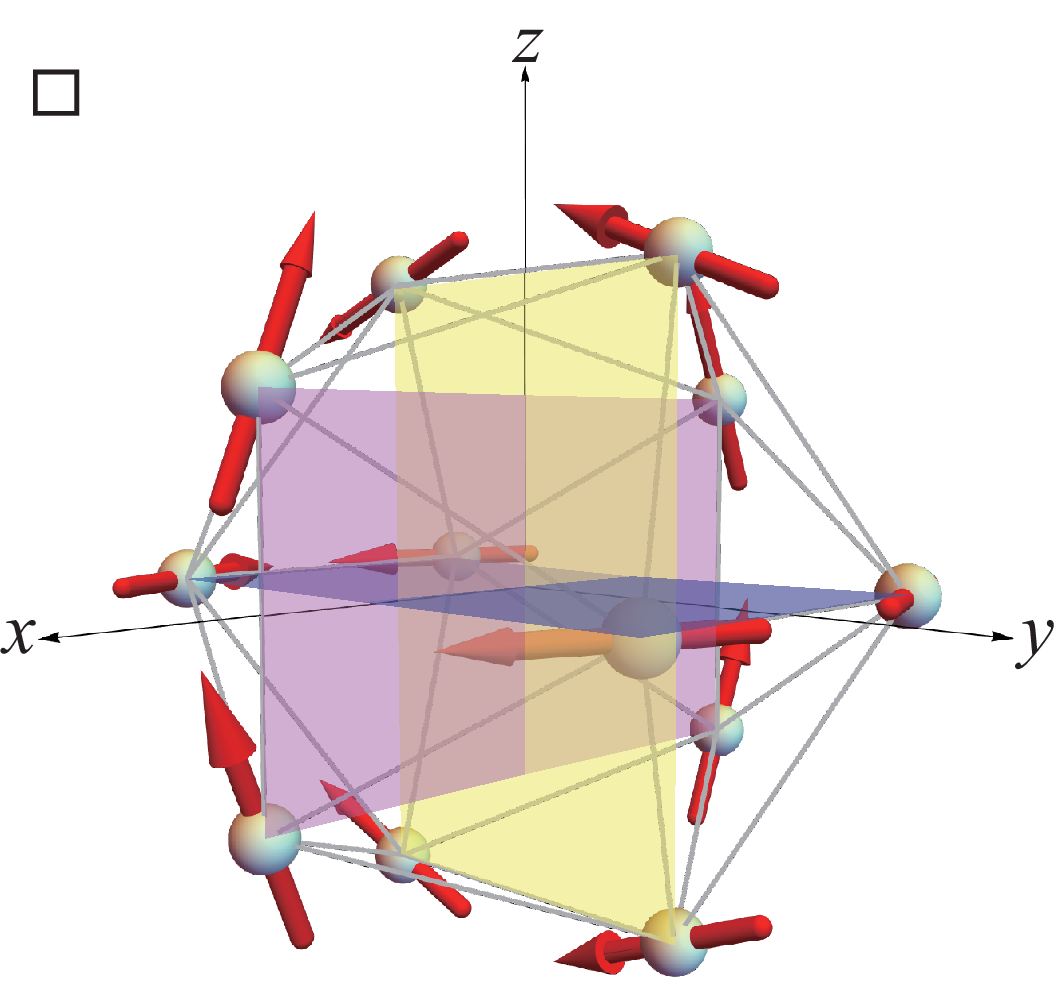}
\caption{(Color online) The non-collinear ferrimagnetic state for $\theta=80^{\circ}$ on the icosahedron denoted by the black-square symbol in Fig.~\ref{fig:uniform}(d).}
\label{fig:black_sq_IC}
\end{figure}

The magnetic state shown in Fig.~\ref{fig:uniform}(c) gives the total scalar chirality ${\bm \chi}^{\rm T}=(-0.323,-0.323,-0.323)$. 
The absolute value of each component of ${\bm \chi}^{\rm T}$ is the same as 
$|\chi^{\rm T}_x|=|\chi^{\rm T}_y|=|\chi^{\rm T}_z|$ for this state irrespective of the anisotropy angle $\theta$. 
The anisotropy-angle $\theta$ dependence of $|\chi^{\rm T}_{\mu}|~(\mu=x,y,z)$ is shown in Fig.~\ref{fig:chi_T_x}. 
We see the non-monotonic $\theta$ dependence where the magnitude vanishes at $\theta_0\equiv\tan^{-1}(\tau)=58.28254\dots^{\circ}$ where 
$\tau$ is the golden mean $\tau=(1+\sqrt{5})/2$. 
When $\theta=\theta_0$, the magnetic moments on the N.N. sites are aligned parallelly as shown in Fig.~\ref{fig:ferri_IC}(b) so that the scalar chirality $\chi_{ijk}$ defined in Eq.~(\ref{eq:SC}) at the triangle $ijk$ sites on the icosahedron becomes zero, giving rise to ${\bm \chi}^{\rm T}={\bf 0}$. 
In Fig.~\ref{fig:chi_T_x}, the $\theta$ dependence of the absolute value of each component of the total magnetic moment of the icosahedron $|J_{{\rm IC}, \mu}|$ is shown (right axis). 
As $\theta$ increases, $|J_{{\rm IC},\mu}|$ increases and has a maximum at $\theta=\theta_0$. Then $|J_{{\rm IC},\mu}|$ decreases as $\theta$ increases from $\theta_0$. 
The topological charge of this state is zero $n=0$. 
These characteristics are summarized in Table~\ref{tb:mag_state}.

The black-square symbol in Fig.~\ref{fig:PD} represents the ferrimagnetic order of the non-collinear ferrimagnetic state shown in Fig.~\ref{fig:uniform}(d).
Figure~\ref{fig:black_sq_IC} illustrates the magnetic structure on the icosahedron for $\theta=80^{\circ}$. 

The magnetic space group of this state is $C2'/m'$ (No. 12.5.70 in the OG notation and No. 12.62 $C2'/m'$ in the BNS notation) [50,51]. Namely, the alignment of the magnetic moments of this state is invariant under the 180$^{\circ}$ rotation around the $x$ axis with inversion of the magnetic moment. This state is also invariant under the mirror operation with respect to the mirror plane perpendicular to the $x$ axis. 

\begin{figure}
\includegraphics[width=8cm]{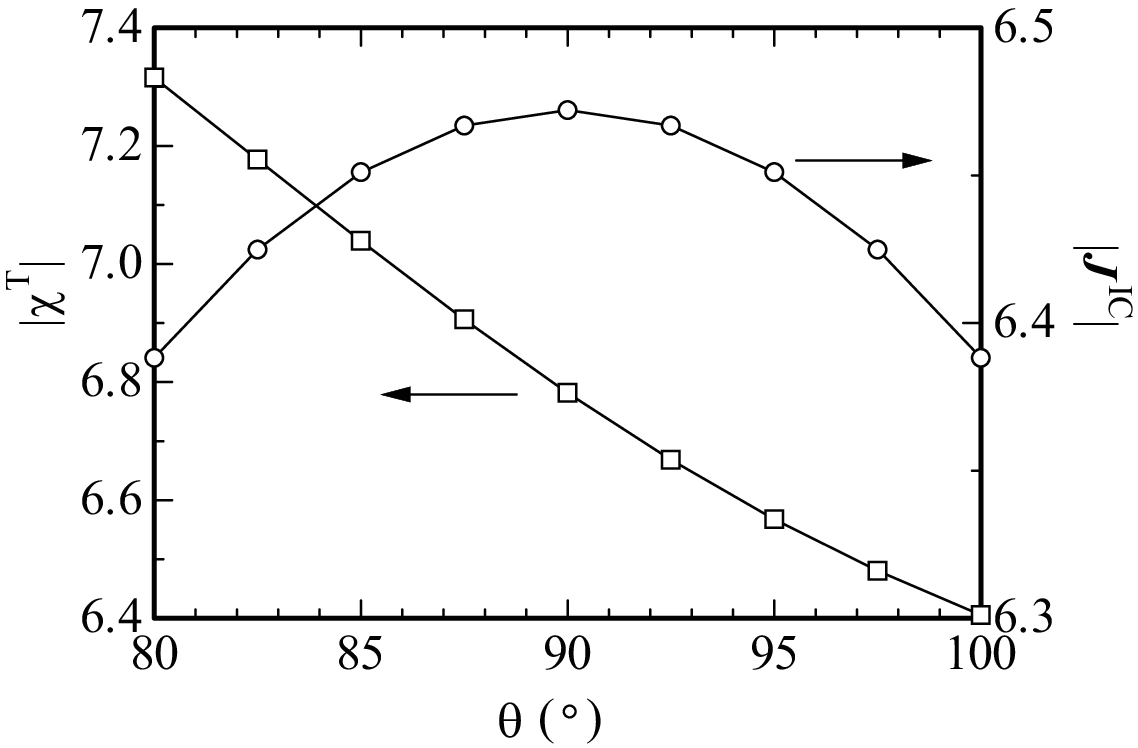}
\caption{The anisotropy-angle $\theta$ dependence of $|\chi^{\rm T}|$ (left axis) and $|{\bm J}_{\rm IC}|$ (right axis) in the magnetic state shown in Fig.~\ref{fig:uniform}(d). 
}
\label{fig:chi_T_black_square}
\end{figure}

\begin{figure}
\includegraphics[width=6cm]{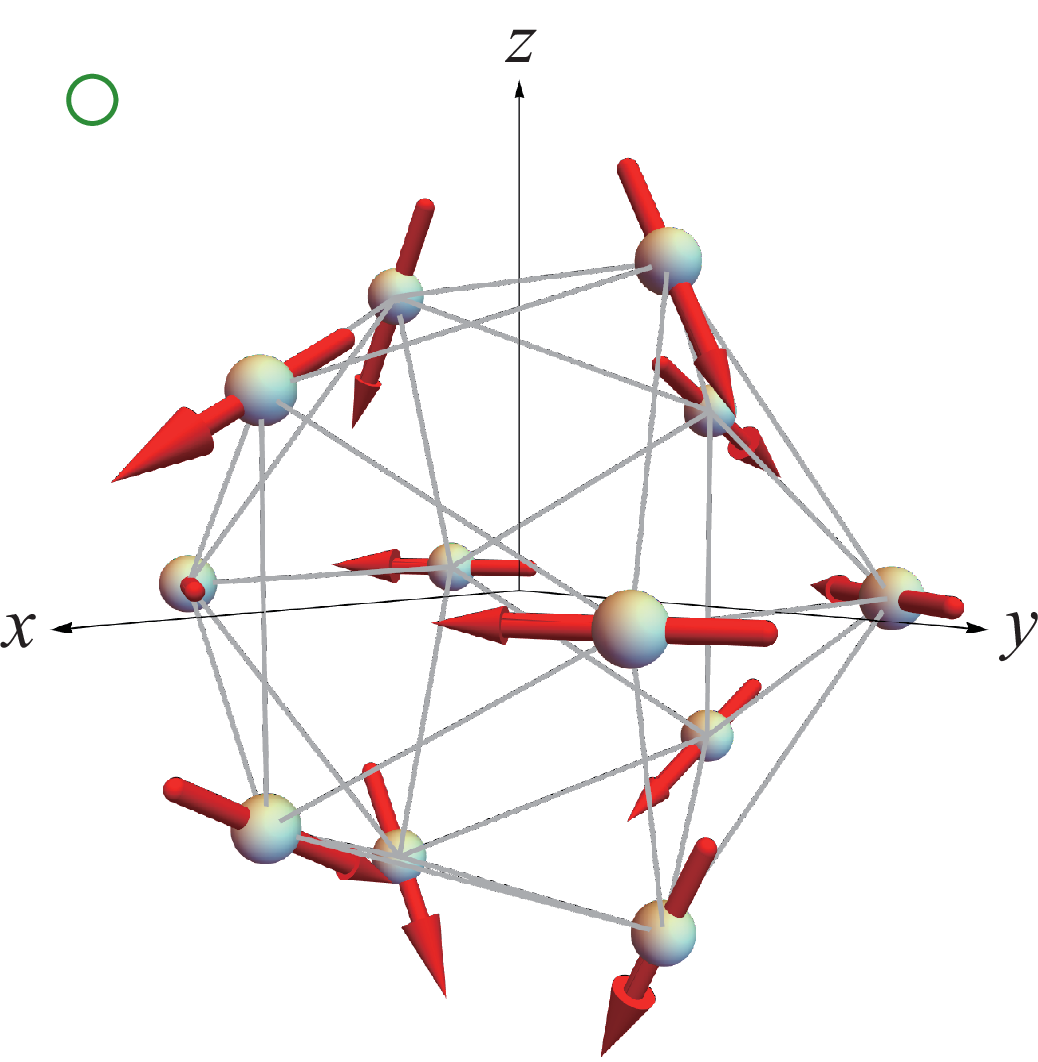}
\caption{(Color online) The non-collinear ferrimagnetic state for $\theta=120^{\circ}$ on the icosahedron denoted by the green-circle symbol in Fig.~\ref{fig:uniform}(e).}
\label{fig:green_circle_IC}
\end{figure}

The magnetic state shown in Fig.~\ref{fig:black_sq_IC} has the net magnetic moment of the icosahedron ${\bm J}_{\rm IC}=(0.00,-5.29,3.72)$ and the total chirality ${\bm \chi}^{\rm T}=(0.00,-2.04,-7.03)$.
The anisotropy-angle $\theta$ dependence of $|\chi^{\rm T}|$ and $|{\bm J}_{\rm IC}|$ is shown in Fig.~\ref{fig:chi_T_black_square}. 
As $\theta$ increases, $|\chi^{\rm T}|$ decreases monotonically while $|{\bm J}_{\rm IC}|$ has a maximum at $\theta=90^{\circ}$. 
The topological charge of this state is zero $n=0$.
There exist the 12 degenerated states with this state which have the same energy. 
These features are summarized in Table~\ref{tb:mag_state}. 


The green-circle symbol in Fig.~\ref{fig:PD} represents the non-collinear ferrimagnetic state shown in Fig.~\ref{fig:uniform}(f).
Figure~\ref{fig:green_circle_IC} shows the magnetic structure on the icosahedron for $\theta=120^{\circ}$. 

The magnetic space group of this state is $C2'/m'$ (No. 12.5.70 in the OG notation and No. 12.62 $C2'/m'$ in the BNS notation) [50,51]. Namely, the alignment of the magnetic moments of this state is invariant under the 180$^{\circ}$ rotation around the $x$ axis with inversion of the magnetic moment. This state is also invariant under the mirror operation with respect to the mirror plane perpendicular to the $x$ axis. 

The magnetic state shown in Fig.~\ref{fig:green_circle_IC} has the net magnetic moment of the icosahedron ${\bm J}_{\rm IC}=(0.00,-3.52,-5.42)$ and the total chirality ${\bm \chi}^{\rm T}=(0.00,1.67,0.00)$.  The topological charge of this state is zero $n=0$. These features are summarized in Table~\ref{tb:mag_state}. 

\begin{figure}
\includegraphics[width=8cm]{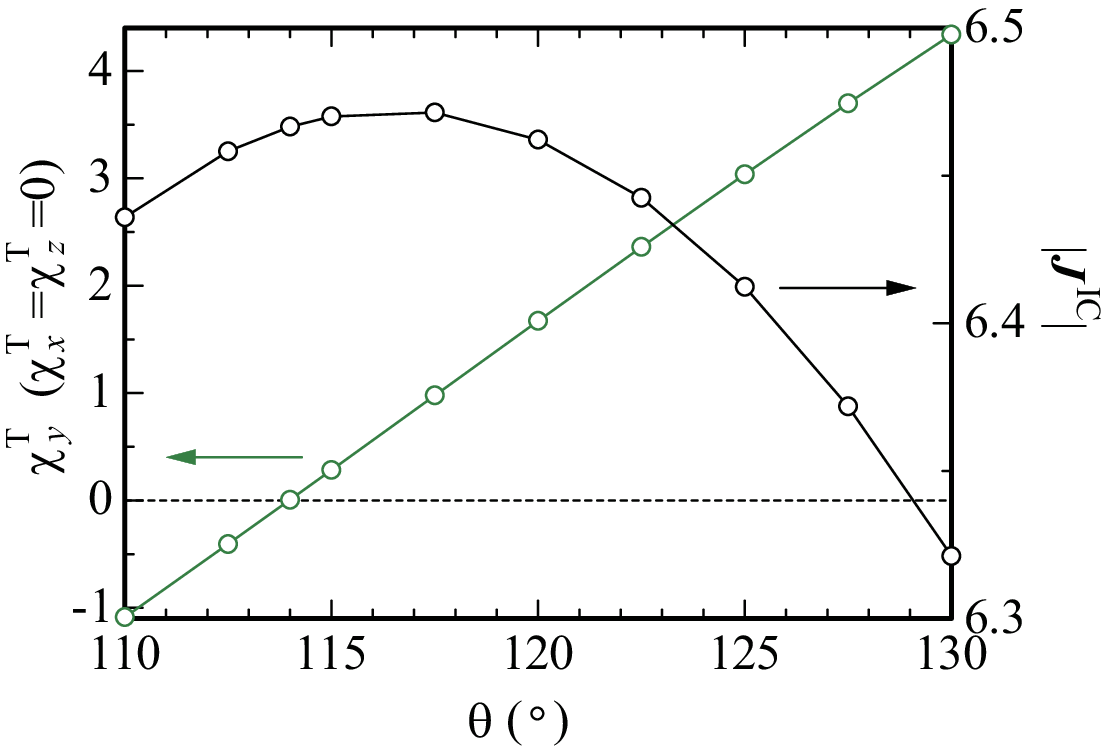}
\caption{(Color online) The anisotropy-angle $\theta$ dependence of $\chi^{\rm T}_{y}$ (left axis) and $|{\bm J}_{\rm IC}|$ (right axis) in the magnetic state shown in Fig.~\ref{fig:uniform}(f). 
The horizontal dashed line indicates $\chi^{\rm T}_{\mu}=0~(\mu=x, y, z)$.}
\label{fig:chi_T_gc}
\end{figure}

\begin{figure}
\includegraphics[width=8cm]{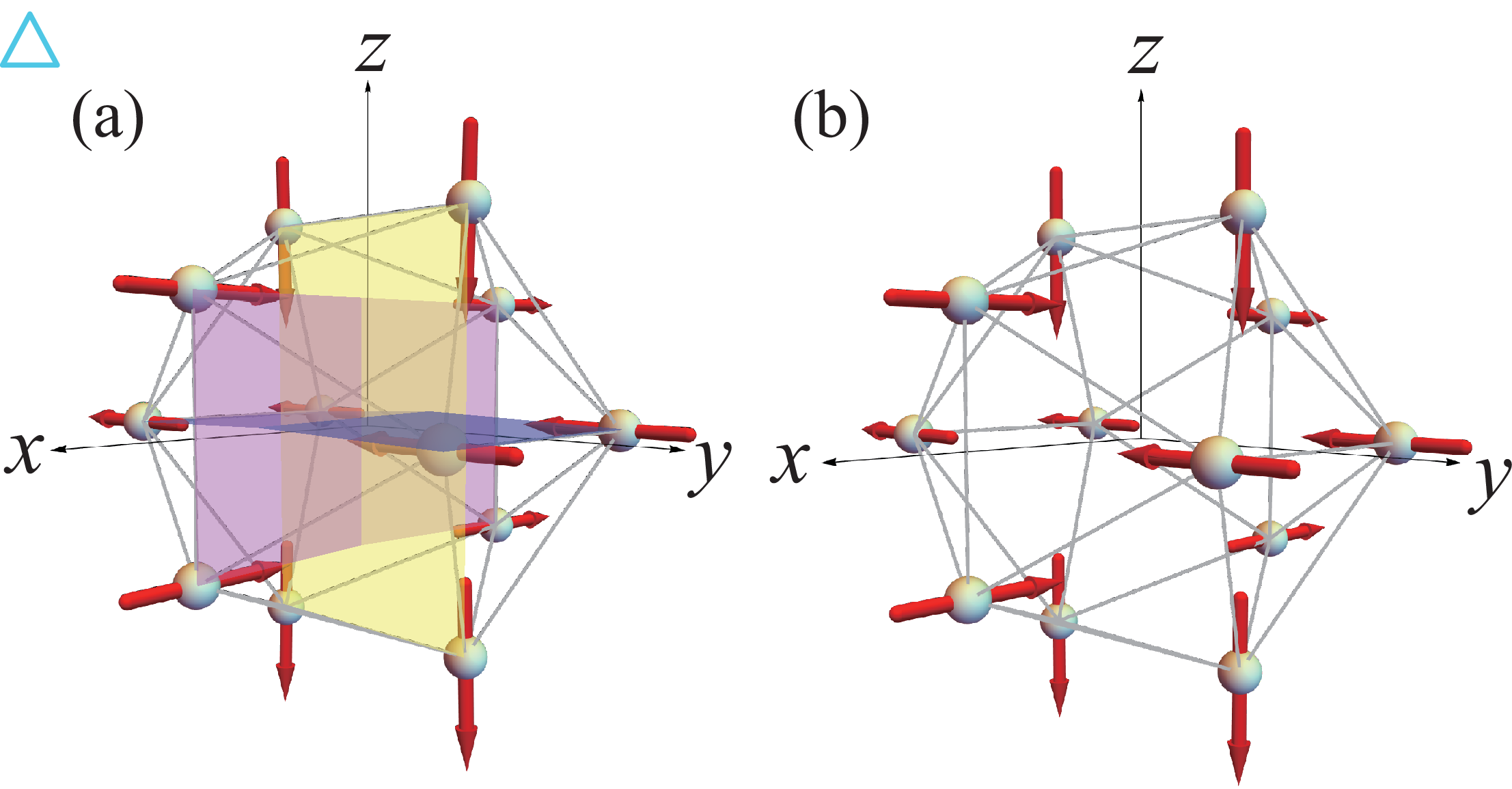}
\caption{(Color online) The non-collinear ferrimagnetic states for (a) $\theta=150^{\circ}$ and (b) $\theta=148.28254^{\circ}$ on the icosahedron denoted by the light-blue-triangle symbol in Fig.~\ref{fig:uniform}(e).}
\label{fig:light_blue_tri_IC}
\end{figure}

The anisotropy-angle $\theta$ dependence of $\chi^{\rm T}_y$ and $|{\bm J}_{\rm IC}|$ is shown in Fig.~\ref{fig:chi_T_gc}. 
As $\theta$ increases, $\chi^{\rm T}_y$ increases monotonically with sign change at $\theta=114^{\circ}$. 
The vanishment of $\chi^{\rm T}_{y}$ occurs at $\theta=114^{\circ}$. 
This is caused by the cancellation of the positive and negative contributions as the result of 
the summation of the finite scalar chirality $\chi_{ijk}$ at each trianglar surface of the icosahedron in Eq.~(\ref{eq:chi_T}). 
This is in sharp cotrast to the $\theta=\theta_0$ case shown in Fig.~\ref{fig:chi_T_x} where each scalar chirality vanishes because of the parallel alignment of the N.N. magnetic moments [see Fig.~\ref{fig:ferri_IC}(b)]. 


The light-blue-triangle symbol in Fig.~\ref{fig:PD} represents the ferrimagnetic order of the non-collinear ferrimagnetic state shown in Fig.~\ref{fig:uniform}(g). 
The 3-fold symmetry with the inversion center with respect to the atomic position holds along the (111) direction. Hence, similarly to the state denoted by the pink-inverted-triangle symbol, the magnetic space group of this state is $R\bar{3}$ (No. 148.1.1247 in OG notation and 148.17 $R\bar{3}$ in BNS notation) [50,51].

Figure~\ref{fig:light_blue_tri_IC}(a) illustrates the magnetic structure on the icosahedron for $\theta=150^{\circ}$.
The net magnetic moment of the icosahedron is directed as ${\bm J}_{\rm IC}=(-4.00,-4.00,-4.00)$ and the total chirality is given by ${\bm \chi}^{\rm T}=(-0.24,-0.24,-0.24)$. 


\begin{figure}[h]
\includegraphics[width=8cm]{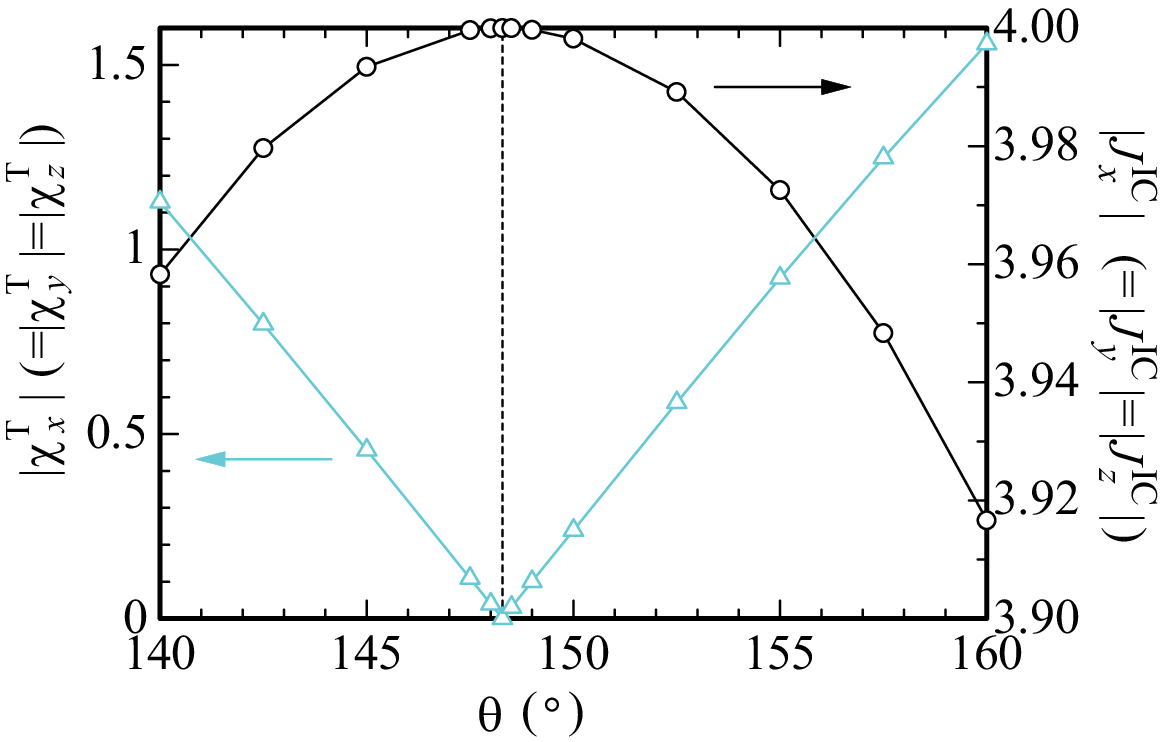}
\caption{(Color online) The anisotropy-angle $\theta$ dependence of $|\chi^{\rm T}_{\mu}|~(\mu=x, y, z)$ (left axis) and the component of $|J_{{\rm IC}, \mu}|~(\mu=x, y, z)$ in the magnetic state shown in Fig.~\ref{fig:uniform}(g).
The vertical dashed line indicates $\theta=\theta_0+90^{\circ}$ (see text).}
\label{fig:chi_T_lbt}
\end{figure}

The anisotropy-angle $\theta$ dependence of each component of the total chirality $|\chi^{\rm T}_{x}| (=|\chi^{\rm T}_y|=|\chi^{\rm T}_z|)$ is shown in Fig.~\ref{fig:chi_T_lbt}. 
The magnitude of each component of the total chirality shows non-monotonic $\theta$ dependence. 
At $\theta=\theta_0+90^{\circ}=148.28254\dots^{\circ}$, $|\chi^{\rm T}_{\mu}|~(\mu=x,y,z)$ becomes zero. 
When $\theta=\theta_0+90^{\circ}$, the magnetic moments at the N.N. sites are aligned parallelly, which are perpendicular to the N.N. bond direction as shown in Fig.~\ref{fig:light_blue_tri_IC}(b). 
Then the scalar chirality $\chi_{ijk}$ defined by Eq.~(\ref{eq:SC}) at the triangle $ijk$ sites on the icosahedron becomes zero, giving rise to ${\bm \chi}^{\rm T}={\bf 0}$. 
The topological charge of this state is zero $n=0$. 
There exist 8 degenerated states with this state which have the same energy. 
These characteristics are summarized in Table~\ref{tb:mag_state}. 
It is noted that at $J_2/J_1=0$, we confirmed that the degeneracy is 192 for $\theta=110^{\circ}$ and $120^{\circ}$, and 64 for $\theta=130^{\circ}$ and $140^{\circ}$ where the magnetic states with the degeneracy 8 listed in Table~\ref{tb:mag_state} are included.

\subsection{Effect of a magnetic field}

In this section, we discuss the effect of a magnetic field. 
We consider the effective model under a magnetic field ${\bm H}$ 
\begin{eqnarray}
{\cal H}=-\sum_{\langle i,j\rangle}J_{ij}\hat{\bm J}_i\cdot\hat{\bm J}_j-\sum_{i}\hat{\bm J}_i\cdot{\bm H}, 
\label{eq:Hh}
\end{eqnarray}
where the first term is the effective model Eq.~(\ref{eq:H}) and the second term is the Zeeman term. 

The total magnetization in the unit cell of the bcc lattice, which is defined as the quantity per icosahedron, is given by
\begin{eqnarray}
{\bm J}_{\rm AC}=\frac{1}{2}\left({\bm J}_{\rm center \ IC}+{\bm J}_{\rm corner \ IC}\right), 
\label{eq:J_uc}
\end{eqnarray}
where the summation of the total magnetization defined by Eq.~(\ref{eq:J_IC}) at the center icosahedron ${\bm J}_{\rm center \ IC}$ and the corner icosahedron ${\bm J}_{\rm corner \ IC}$ in the unit cell of the bcc lattice 
are taken. 

The total chirality in the unit cell which is defined as the quantity per icosahedron is given by 
\begin{eqnarray}
{\bm \chi}^{\rm T}=\frac{1}{2}\left({\bm \chi}^{\rm T}_{\rm center \ IC}+{\bm \chi}^{\rm T}_{\rm corner \ IC}\right), 
\label{eq:chi_T_uc}
\end{eqnarray}
where the summation of the total chirality defined by Eq.~(\ref{eq:chi_T}) at the center icosahedron ${\bm \chi}^{\rm T}_{\rm center \ IC}$ and the corner icosahedron ${\bm \chi}^{\rm T}_{\rm corner \ IC}$ in the unit cell of the bcc lattice 
are taken. 

We have analyzed the effects of the magnetic field on the hedgehog-anti-hedgehog order and the whirling-anti-whirling order. 
It is noted that the main result was reported in ref.~\cite{WPNAS} and here the detailed analyzes are given. 

\subsubsection{Hedgehog-anti-hedgehog state}

\begin{figure}[tb]
\includegraphics[width=8cm]{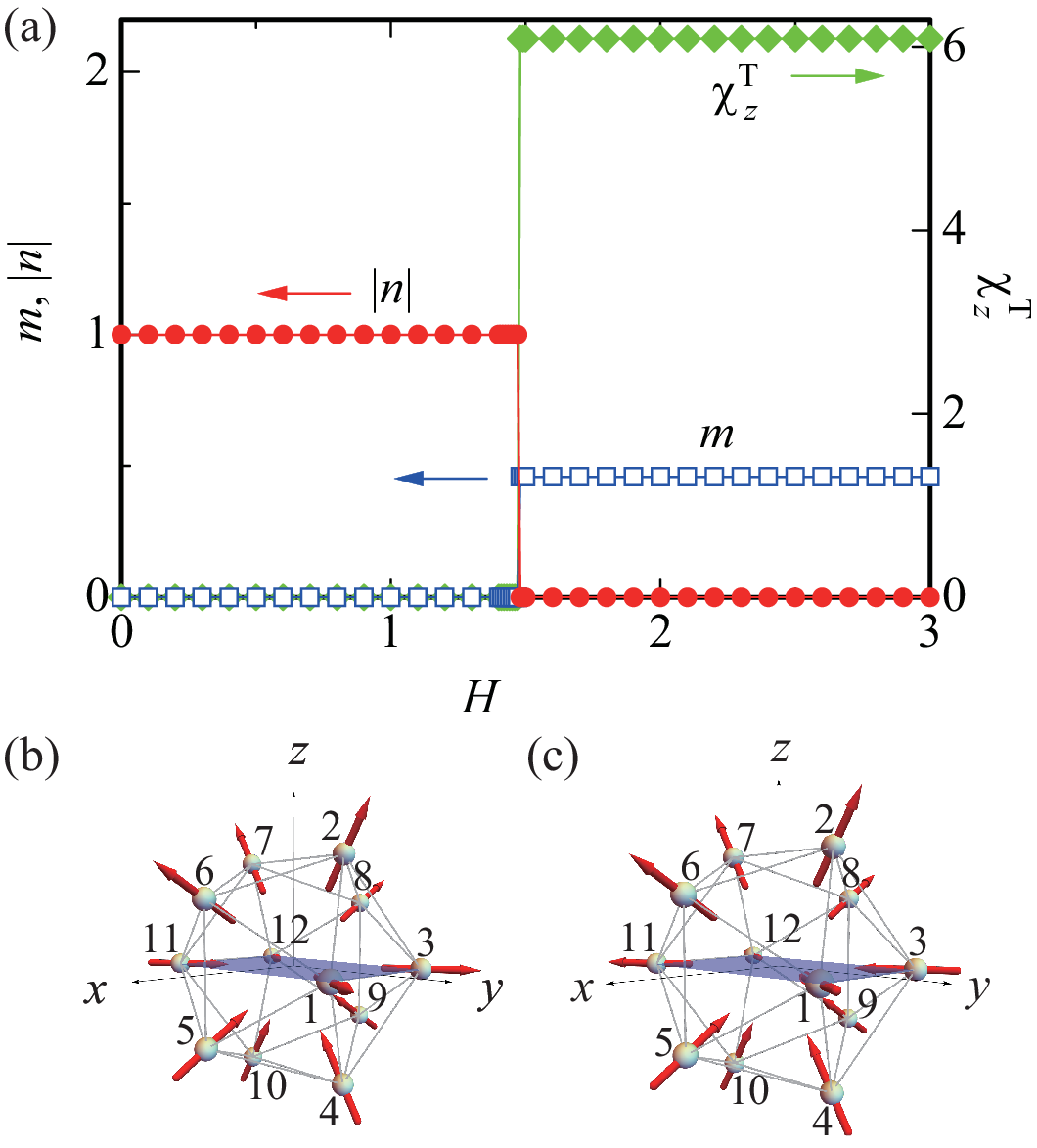}
\caption{(Color online) 
(a) The magnetic field dependence of the magnetization $m$, the topological charge $|n|$, and the total chirality $\chi^{\rm T}_z$ for $J_1=1$ and $J_2=0.1$ at $\theta=0^{\circ}$. 
(b) Magnetic states at the center and corner icosahedrons in the unit cell of the bcc lattice for $H>H_{\rm M}$.
(c) The energetically degenerate state with the magnetic state in (b) for $H>H_{\rm M}$. 
In (b) and (c), the number labels the rare-earth site on the icosahedron.
}
\label{fig:J_h_HAH}
\end{figure}

We apply the magnetic field ${\bm H}=(0,0,H)$ to the hedgehog-anti-hedgehog order shown in Fig.~\ref{fig:H_AH_W_AW}(a). 
We show the result for $J_1=1$ and $J_2=0.1$ at the anisotropy angle $\theta=0^{\circ}$ as a representative case. 
The field dependence of the magnetization per the rare-earth atom 
\begin{eqnarray}
m=\frac{J_{{\rm AC},z}}{12} 
\label{eq:mag}
\end{eqnarray}
is shown in Fig.~\ref{fig:J_h_HAH}(a) (left axis).
The magnetization suddenly shows a jump at $H=H_{\rm M}\equiv 1.48$. This metamagnetic transition occurs by the flip of 
the half magnetic moments among the 12 moments of the hedgehog state and the anti-hedgehog state on each icosahedron in the unit cell, which results in the magnetic state shown in Fig.~\ref{fig:J_h_HAH}(b). 
This state is uniformly distributed at the center and corner icosahedrons in the unit cell for $H>H_{\rm M}$. 
For $H\ge H_{\rm M}$, the magnetic state shown in Fig.~\ref{fig:J_h_HAH}(c) is also energetically degenerate to the state shown in Fig.~\ref{fig:J_h_HAH}(b). 
Both the states are created by inverting 
the magnetic moments at the $i=4$, 5, 9, 10 site from the hedgehog state or the magnetic moments at the $i=2$, 6, 7, 8 site from the anti-hedgehog state. 
Among the 4 magnetic moments at each vertex of the purple plane which is perpendicular to the magnetic field, the magnetic moments at the $i=11$ and $12$ ($i=1$ and 3) sites are flipped in Fig.~\ref{fig:J_h_HAH}(b) 
and the magnetic moments at the $i=1$ and $3$ ($i=11$ and 12) sites are flipped in Fig.~\ref{fig:J_h_HAH}(c) from the hedgehog state (anti-hedgehog state). 
The topological charge of both the states on the icosahedron is zero $n=0$. 
The characteristics are summarized in Table~\ref{tb:h}.

For $H<H_{\rm M}$, the hedgehog state with $n=+1$ and the anti-hedgehog state with $n=-1$ are realized on each icosahedron in the unit cell. 
At $H=H_{\rm M}$, the metamagnetic transition takes place simultaneously with the topological transition from $|n|=1$ to $n=0$ [see Fig.~\ref{fig:J_h_HAH}(a) (left axis)].
For $H>H_{\rm M}$, 
both the states shown in Figs.\ref{fig:J_h_HAH}(b) and \ref{fig:J_h_HAH}(c) have the common $z$ component of the magnetization and the total chirality with the $x$ component being zero and have the $y$ component with opposite sign (see Table~\ref{tb:h}). 
This implies that the topological Hall effect can be observed after the metamagnetic transition for $H>H_{\rm M}$ by measuring the Hall conductivity $\sigma_{xy}$ via Eq.~(\ref{eq:sigma}). 
Figure~\ref{fig:J_h_HAH}(a) shows the magnetic-field dependence of $\chi^{\rm T}_{z}$ (right axis). We see that $\chi^{\rm T}_z$ changes from $0$ to finite at the metamagnetic transition $H=H_{\rm M}$.

\begin{table*}
\caption{
Characteristics of the magnetic state for $H>H_{\rm M}$ in the 1/1 AC. 
As for the distribution, u indicates the uniform distribution of the magnetic states on the center icosahedron and the corner icosahedron in the bcc unit cell of the 1/1 AC. 
}
\label{tb:h}
\begin{tabular}{lccccc}
\hline
\hline
\textrm{magnetic states}&
\textrm{distribution}&
\textrm{degeneracy}& \textrm{${\bm J}_{\rm AC}$} & \textrm{${\bm \chi}^{\rm T}$} &
{\textrm{topological charge}}\\
\hline
Fig.~\ref{fig:J_h_HAH}(b) & u & 2 & $(0.00, 3.40,5.51)$ & $(0.00, 3.04,6.09)^{\rm a}$ & 0 \\
Fig.~\ref{fig:J_h_HAH}(c) & & & $(0.00,-3.40,5.51)$ & $(0.00,-3.04,6.09)$ & 0 \\
\hline
Fig.~\ref{fig:J_h_WAW}(b) & u & 2 & $(3.40, 0.00, 5.51)$ & $(-5.64, 0.00, 3.76)^{\rm b}$ & 0 \\
Fig.~\ref{fig:J_h_WAW}(c) & & & $(-3.40, 0.00, 5.51)$ & $(5.64, 0.00, 3.76)$ & 0   
\\
\hline
\hline
\end{tabular}
\\
$^a$ \footnotesize{$\theta=0^{\circ}$ case.}
\\
$^b$ \footnotesize{$\theta=90^{\circ}$ case.}
\end{table*}

\subsubsection{Whirling-anti-whirling state}

\begin{figure}[tb]
\includegraphics[width=8cm]{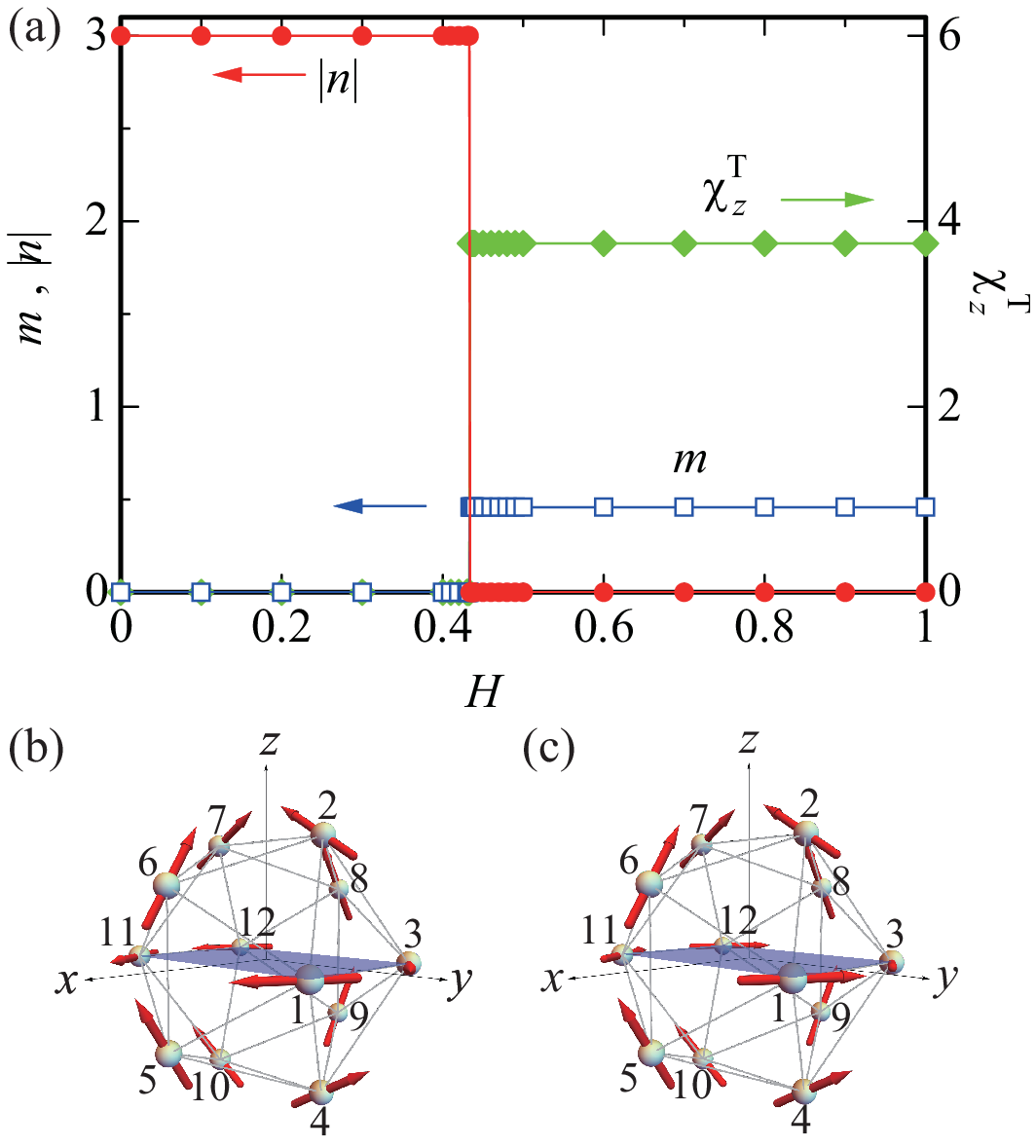}
\caption{(Color online) 
(a) The magnetic field dependence of the magnetization $m$, the topological charge $|n|$, and the total chirality $\chi^{\rm T}_z$ for $J_1=1$ and $J_2=8$ at $\theta=90^{\circ}$. 
(b) Magnetic state at the center and corner icosahedrons in the unit cell of the bcc lattice for $H>H_{\rm M}$.
(c) The energetically degenerate state with the magnetic state in (b) for $H>H_{\rm M}$. 
In (b) and (c), the number labels the rare-earth site on the icosahedron.
}
\label{fig:J_h_WAW}
\end{figure}

We apply the magnetic field ${\bm H}=(0,0,H)$ to the whirling-anti-whirling order illustrated in Fig.~\ref{fig:H_AH_W_AW}(b). 
We show the result for $J_1=1$ and $J_2=8$ at the anisotropy angle $\theta=90^{\circ}$ as a representative case. 
The magnetic-field dependence of the magnetization $m$ defined by Eq.~(\ref{eq:mag}) is shown in Fig.~\ref{fig:J_h_WAW}(a) (left axis). 
The magnetization $m$ exhibits an abrupt jump at $H=H_{\rm M}=0.44$, which indicates the metamagnetic transition. 
For $H>H_{\rm M}$, the half of the 12 magnetic moments of the whirling-moment state and anti-whirling-moment state on each icosahedron is inverted by the magnetic field, which gives rise to the magnetic state shown in Fig.~\ref{fig:J_h_WAW}(b). This state is uniformly distributed at the center icosahedron and the corner icosahedron in the bcc unit cell. 
The magnetic state shown in Fig.~\ref{fig:J_h_WAW}(c) is also energetically degenerate to the state shown in Fig.~\ref{fig:J_h_WAW}(b). 
Both the states can be created by inverting the magnetic moments at the $i=2$, 5, 7, 9 site from the anti-whirling-moment state or the magnetic moments at the $i=4$, 6, 8, 10 site from the whirling-moment state. 
Among the 4 magnetic moments at each vertex of the purple plane in Figs.~\ref{fig:J_h_WAW}(b) and \ref{fig:J_h_WAW}(c) which is perpendicular to the magnetic field, the magnetic moments at the $i=3$ and $12$ ($i=1$ and 11) sites are flipped in Fig.~\ref{fig:J_h_WAW}(b) 
and the magnetic moments at the $i=1$ and $11$ ($i=3$ and 12) sites are flipped in Fig.~\ref{fig:J_h_WAW}(c) from the whirling-moment state (anti-whirling-moment state). 
The topological charge of both the states on the icosahedron is zero $n=0$. 
The characteristics are summarized in Table~\ref{tb:h}. 

For $H<H_{\rm M}$, the whirling-moment state with $n=+3$ and the anti-whirling-moment state with $n=-3$ are realized on each icosahedron in the unit cell. 
At $H=H_{\rm M}$, the metamagnetic transition takes place simultaneously with the topological transition from $|n|=3$ to $n=0$ [see Fig.~\ref{fig:J_h_WAW}(a) (left axis)].
For $H>H_{\rm M}$, 
both the states shown in Figs.~\ref{fig:H_AH_W_AW}(b) and \ref{fig:H_AH_W_AW}(c) have the common $z$ component of the magnetization and total chirality with the $y$ component being zero and have the $x$ component with opposite sign. 
This implies that the topological Hall effect can be observed after the metamagnetic transition for $H>H_{\rm M}$ by measuring the Hall conductivity $\sigma_{xy}$ via Eq.~(\ref{eq:sigma}). 
Figure~\ref{fig:J_h_WAW}(a) shows the magnetic-field dependence of $\chi^{\rm T}_{z}$ (right axis). We see that $\chi^{\rm T}_z$ changes from $0$ to finite at the metamagnetic transition $H=H_{\rm M}$.

Recently, theoretical calculation for the angle dependence from the (001) to the (111) directions for applied magnetic field on the whirling-anti-whirling state has been performed~\cite{WI2026}. By applying a magnetic field along the direction between (001) and (111) for $H>H_{\rm M}$, the degeneracy shown as Figs.~\ref{fig:J_h_WAW}(b) and \ref{fig:J_h_WAW}(c) is lifted due to the low symmetry, which enables us to achieve a single domain. 

\section{Discussion}

On the basis of the effective model (\ref{eq:H}), the ground-state phase diagram has been determined as Fig.~\ref{fig:PD} by numerical calculations. 
The hedgehog-anti-hedgehog state is realized in the vicinity of $\theta=0^{\circ}$ for small $J_2/J_1$ 
and the whirling-anti-whirling state is realized in the vicinity of $\theta=90^{\circ}$ for large $J_2/J_1$. 
The uniform order of the non-collinear ferrimagnetic state denoted by the pink-inverted-triangle symbol is realized in the region of $40^{\circ}\le\theta\le 80^{\circ}$ for the wide range of $J_2/J_1$. 
The emergence of these states were shown to appear in the ground-state phase diagram of the 1/1 AC determined in ref.~\cite{WPNAS}. Hence, these states were correctly pointed out as the ground state in the 1/1 AC in the previous study by the energy comparison based on the magnetic ground state on the icosahedron~\cite{WPNAS}. 
In ref.~\citen{Sato2019}, it was reported that the magnetic moment of the whirling-anti-whirling ordered state in the 1/1 AC Au$_{72}$Al$_{14}$Tb$_{14}$ is lying in the mirror plane [i.e., the $yz$ plane in the local coordinate shown in Fig.~\ref{fig:Tb_local}(f)] with the anisotropy angle $\theta=86^{\circ}$ from the pseudo five-fold axis. 
The measured temperature dependence of the magnetic susceptibility and the magnetic-field dependence of the magnetization are well reproduced by the effective model (\ref{eq:H}) applied to the single icosahedron for large $J_2/J_1$~\cite{Sato2019,Sato_pc}. 

Recently, the similar magnetic structure has been observed in the 1/1 AC Au$_{65}$Ga$_{21}$Tb$_{14}$~\cite{Nawa2023}. 
The measured magnetization curve is shown to be well reproduced by the effective model (\ref{eq:H}) applied to the single icosahedron for $J_2/J_1=5.982$. 
In Fig.~\ref{fig:PD}(a), the whirling-anti-whirling order is realized for $J_2/J_1\ge8.7$ at $\theta=85^{\circ}$. 
The slightly larger $J_2/J_1$ than that reported in ref.~\citen{Nawa2023} is considered to be due to the difference between the single icosahedron analysis in ref.~\citen{Nawa2023} and the analysis in the 1/1 AC in the present study. 

Recently, the whirling-anti-whirling state in the 1/1 AC Au$_{64}$Ga$_{22}$Tb$_{14}$ has been identified by the neutron measurement~\cite{Labib}. The magnetic space group has been reported to be $I_{\rm P}m'\bar{3'}$ (No. 204.5.1534 in the OG notation and No. 201.21 in the BNS notation)~\cite{Labib}.

The non-collinear ferrimagnetic state denoted by the pink-inverted-triangle symbol was actually observed by the neutron measurement in the 1/1 AC Au$_{70}$Si$_{17}$Tb$_{13}$~\cite{Hiroto}. 
In ref.~\citen{Hiroto}, it was reported that the magnetic moment in the non-collinear ferrimagnetic state in the 1/1 AC Au$_{70}$Si$_{17}$Tb$_{13}$ is lying in the mirror plane [i.e., the $yz$ plane in Fig.~\ref{fig:Tb_local}(f)] with the anisotropy angle $\theta=80^{\circ}$. 
As shown in Fig.~\ref{fig:PD}(a), this state appears for $40^{\circ}\le\theta\le80^{\circ}$, which is in accordance with the measurement.

The magnetization was observed under small magnetic field in the 1/1 AC Au$_{70}$Si$_{17}$Tb$_{13}$~\cite{Hiroto}, which is about $3 \mu_{\rm B}$. Since the saturated moment for an isolated Tb$^{3+}$ ion is $g_{J}\mu_{\rm B}J=9 \mu_{\rm B}$, about $1/3$ of saturate moment is estimated to be the ordered moment per Tb. Our result shown in Fig.~\ref{fig:chi_T_x} shows that $J_z^{\rm IC}\approx 4$ for $\theta=70^{\circ}$, which gives about $1/3$ of full polarization per site. Hence, if $\theta$ is close to $70^{\circ}$ in the 1/1 AC Au$_{70}$Si$_{17}$Tb$_{13}$, the value of the magnetization is consistent with the observations.

The similar magnetic structure was observed in the 1/1 AC Au$_{68.6}$Si$_{17.8}$Tb$_{13.6}$ which was referred to as the TAS(0) state in ref.~\citen{Gebresenbut}. 
The magnetic space group $R\bar{3}$ was identified by analyzing the neutron data~\cite{Gebresenbut}. 
In ref.~\citen{Gebresenbut}, the magnetic structure in the 1/1 AC Au-Si-Ho is refereed to as HAS(52) and that in Au-Si-Tb is referred to as TAS(14). 
The former magnetic structure is similar to that in TAS(0). On the other hand, the latter magnetic structure is slightly modulated from that in TAS(0). This was pointed out to be due to the effect of 
the Tb atom located at the center of the icosahedron [see Fig.~\ref{fig:Tb_local}(a)]. 
In the present study, we have analyzed the effective model in the 1/1 AC with the icosahedron in the primitive unit cell where there exists no Tb atom at the cluster center as shown in Fig.~\ref{fig:Tb_local}(a). 
The theoretical analysis of the effective model on the rare-earth sites not only on the 12 vertices of the icosahedron but also at the cluster center is interesting future subject. 

Experimentally, the whirling-anti-whirling state and the ferrimagnetic state were observed in the 1/1 AC Au$_{70}$Si$_{17}$Tb$_{13}$~\cite{Hiroto} and Au$_{72}$Al$_{14}$Tb$_{14}$~\cite{Sato2019} , respectively. The number of valence electrons per atom, which is referred to as e/a ratio, of each material is 1.77 and 1.56, respectively. As theoretically shown in Ref.~\citen{WPNAS}, the key parameter for controlling the CEF is the ratio of the effective valences of ligand ions $\alpha=Z_{\rm SM}/Z_{\rm Au}$ (SM=Si, Al). The nominal valences of Si and Al in metallic compounds are $+4$ and $+1$, respectively. Hence, the $\alpha$ value is expected to be larger in Au$_{70}$Si$_{17}$Tb$_{13}$ than Au$_{72}$Al$_{14}$Tb$_{14}$ although we should keep in mind that the effect of screening by conduction electrons causes deviation from the nominal valence. As shown in Ref.~\citen{WPNAS}, the anisotropy angle $\theta$ decreases as $\alpha$ increases. Namely, as shown in Fig.~\ref{fig:PD}(a), the whirling-anti-whirling state changes to the ferrimagnetic state as $\theta$ decreases from $\theta=90^{\circ}$. This tendency is consistent with the above experimental observations. 

The same tendency was also observed in the 1/1 AC Au$_{68-x}$Ga$_{18+x}$Tb$_{14}$~\cite{Labib}. The nominal valence of Ga is $+3$. As e/a increases by increasing the Ga composition i.e.,$x$ increases with the Au composition decreasing, $\alpha=Z_{\rm Ga}/Z_{\rm Au}$ is expected to increase. As $\alpha$ increases, the anisotropy angle $\theta$ decreases, as shown in Ref.~\citen{WPNAS}. Hence, the whirling-anti-whirling state changes to the ferrimagnetic state as shown in Fig.~\ref{fig:PD}(a). This tendency is consistent with the temperature-e/a phase diagram [Fig.~13(a)] reported in Ref.~\citen{Labib}.

As summarized in Table~\ref{tb:mag_state}, the total chirality ${\bm \chi}^{\rm T}\ne{\bf 0}$ for the ferrimagnetic orders denoted by the open symbols in Fig.~\ref{fig:PD}. This implies that the emergent fictious magnetic field exists in these states. 
In the non-collinear ferrimagnetic state denoted by the pink-inverted-triangle symbol, there exists the total chirality with finite component along each 2-fold-axis direction except for the anisotropy angle $\theta=\theta_0$ (see Fig.~\ref{fig:chi_T_x}). Although as $\theta$ deviates from $\theta_0$, the magnitude of each component $|\chi^{\rm T}_{\mu}|$~$(\mu=x, y, z)$ increases, the value is relatively smaller than those of the other ferrimagnetic ordered states denoted by the open symbols in Fig.~\ref{fig:PD}. 
In the magnetic states denoted by the blue-square symbol and the green-circle symbol in Fig.~\ref{fig:PD}, there exists the total chirality ${\bm \chi}^{\rm T}$ along the 2-fold-axis direction.
In particular, the former state exhibits relatively large magnitude $|{\bm \chi}^{\rm T}|$ (see Fig.~\ref{fig:chi_T_bs}). 

The magnetic state denoted by the black-square symbol has not been identified experimentally to date. 
The magnetically ordered states identified so far are the whirling-anti-whirling state (orange square) and the uniform order of the non-collinear ferrimagnetic state (pink-inverted triangle) in Fig.~\ref{fig:PD} with the anisotropy angle $\theta$ close to $\theta=90^{\circ}$ as reported in refs.~\citen{Sato2019,Hiroto,Gebresenbut,Nawa2023}. Since the magnetic state denoted by the black-square symbol appears in the region around $\theta=90^{\circ}$ in Fig.~\ref{fig:PD}, there exists the possibility that this state is realized in the rare-earth based 1/1 AC. As shown in Fig.~\ref{fig:chi_T_black_square}, this state exhibits large magnitude of the total chirality $|{\bm \chi}^{\rm T}|$. 

The total chirality $|{\bm \chi}^{\rm T}|\ne 0$ in the ferrimagnetic ordered states denoted by the open symbol in Fig.~\ref{fig:PD} 
is expected to give 
rise to the topological Hall conductivity as Eq.~(\ref{eq:sigma}). 
For example, in the non-collinear ferrimagnetic state denoted by the pink-inverted triangle in Fig.~\ref{fig:PD}, there is the 8 degeneracy as shown in Table~\ref{tb:mag_state}. When the magnetic field is applied to the (111) direction, the degeneracy is lifted so that the unique ground state with ${\bm J}_{\rm IC}\parallel$~(111) and ${\bm \chi}^{\rm T}\parallel$~(-1-1-1) is realized. 
Hence, the topological Hall effect 
is expected to appear  
in the Hall conductivity $\sigma_{xy}$, $\sigma_{yz}$, and $\sigma_{zx}$ by Eq.~(\ref{eq:sigma}). 
In this way, by applying the magnetic filed, the degeneracy of the ground state can be lifted. Then, by measuring the Hall conductivity in the single domain, the topological Hall effect can be detected. Hence, Table~\ref{tb:mag_state} is useful to perform such a measurement.

Recently, the CEF in the QC Au-SM-Dy and the AC has been analyzed theoretically on the basis of the point charge model~\cite{WI2023}. 
Since the Dy$^{3+}$ ion has the $4f^9$ electron configuration, the CEF ground state has the Kramers degeneracy. This is in sharp contrast to the Tb-based systems since the Tb$^{3+}$ ion with the $4f^8$ electron configuration has in principle no Kramers degeneracy. However, the magnetic easy axis for the CEF ground state in the Au-SM-Dy systems has been revealed to lie in the mirror plane [i.e., the $yz$ plane in Fig.~\ref{fig:Tb_local}(f)] and the $\alpha=Z_{\rm SM}/Z_{\rm Au}$ dependence of the magnetic easy axis behaves similarly to that in the Au-SM-Tb systems~\cite{WI2023}. 
This suggests that the model (\ref{eq:H}) can be applied to the broad range of the rare-earth based QCs and ACs with the uniaxial anisotropy arising from the CEF not only for the Tb-based systems but also the other rare-earth-based systems. 

\section{Conclusion}

On the basis of the effective model taking into account the uniaxial anisotropy arising from the CEF, we have determined the ground-state phase diagram in the 1/1 AC 
by performing numerically-exact calculation. 
The hedgehog-anti-hedgehog state is shown to be stabilized near $\theta=0^{\circ}$ and $\theta=180^{\circ}$ for small $J_2/J_1$ 
and the whirling-anti-whirling state is realized in the vicinity of $\theta=90^{\circ}$ for large $J_2/J_1$ in the phase diagram. 
These antiferromagnetic ordered states have no net magnetic moment on the icosahedron which is shown to be characterized by the topological charge as the hedgehog-anti-hedgehog order with $|n|=1$ and the whirling-anti-whirling order with $|n|=3$. 
Although the total chirality in these states is zero, by applying a magnetic field, the metamagnetic transition occurs at $H=H_{\rm M}$ simultaneously with the topological transition where the finite total chirality emerges for $H>H_{\rm M}$. Hence, the topological Hall effect is expected to appear after the metamagnetic transition. 

In the uniform orders of the non-collinear ferrimagnetic state, the net magnetization on the icosahedron is not zero with topological charge being zero $n=0$ and the total chirality is finite. 
At zero field, the non-collinear ferrimagnetic state is energetically degenerate. By applying a magnetic field, the degeneracy can be lifted where the topological Hall effect can be observed. 

Our results are shown to explain the magnetic structures observed in the 1/1 AC Au$_{72}$Al$_{14}$Tb$_{14}$~\cite{Sato2019}, Au$_{65}$Ga$_{21}$Tb$_{14}$~\cite{Nawa2023} and Au$_{70}$Si$_{17}$Tb$_{13}$~\cite{Hiroto} as well as Au$_{68.6}$Si$_{17.8}$Tb$_{13.6}$ and Au-Si-Ho referred to as HAS(52) in ref.~\citen{Gebresenbut}. 
Namely, we have succeeded in explaining the whilirling-antiwhirling state and noncollinear ferrimagnetic state identified experimentally to date. This indicates that our minimum model (1) captures the essence of magnetism in the rare-earth based 1/1 ACs, which is regarded as the effective model. 
The magnetic states revealed in the present study are expected to be useful to understand the 1/1 AC Au-SM-R and Cd$_6$R systems whose magnetic structures have not been identified experimentally. 
We have clarified the degeneracy of the ground state and reveled each property. This is considered to be important in understanding the domain structures of the magnetically ordered states, which inevitably exist in real materials. 
The effective model is expected to be applied to the broad range of the rare-earth based ACs and QCs with the uniaxial anisotropy arising from the CEF.

\begin{acknowledgment}

The author acknowledges K. Momma for 
expanding functionality of the software VESTA~\cite{Momma}
which helps to analyze the magnetic structure in the 1/1 AC as well as kind instructions. 
The author thanks valuable discussions about the magnetic space group with K. Momma, K. Nawa, F. Labib, and R. Tamura, H. Takakura and M. de Boissieu. 
This work was supported by JSPS KAKENHI Grant Numbers JP18K03542, JP19H00648, JP22H04597, JP22H01170, JP23K17672, and JP24H01675.

\end{acknowledgment}

* swata@mns.kyutech.ac.jp

\end{document}